\documentclass[12pt,a4paper]{article}
\usepackage{type1cm,amsmath,hangcaption,graphicx,indentfirst}
\usepackage[psamsfonts]{amssymb}
\numberwithin{equation}{section}

\DeclareMathOperator*{\Tr}{{\rm Tr}}

\newcommand{\rs}{{ \mathbb R} \times S^3/{{\mathbb Z}_k}}
\newcommand{\sz}{S^3/{{\mathbb Z}_k}}

\begin{document}
\begin{titlepage}

 \renewcommand{\thefootnote}{\fnsymbol{footnote}}
\begin{flushright}
 \begin{tabular}{l}
 DESY 06-179\\
 hep-th/0610119\\
 \end{tabular}
\end{flushright}

 \vfill
 \begin{center}
 \font\titlerm=cmr10 scaled\magstep4
 \font\titlei=cmmi10 scaled\magstep4
 \font\titleis=cmmi7 scaled\magstep4
 \centerline{\titlerm Phase Transitions of Large {\titlei N} Orbifold
                      Gauge Theories}
 \vskip 2.5 truecm

\noindent{ \large Yasuaki Hikida\footnote{E-mail:
yasuaki.hikida@desy.de}}
\bigskip

 \vskip .6 truecm
\centerline{\it DESY Theory Group, Notkestrasse 85, D-22603 Hamburg, Germany}

 \vskip .4 truecm

 \end{center}

 \vfill
\vskip 0.5 truecm

\begin{abstract}

We study the phase structures of 
${\cal N}=4$ $U(N)$ super Yang-Mills theories 
on ${\mathbb R} \times S^3/{\mathbb Z}_k$ with large $N$.
The theory has many vacua labelled by
the holonomy matrix along the non-trivial cycle on 
$S^3/{\mathbb Z}_k$, and for the fermions 
the periodic and the anti-periodic boundary 
conditions can be assigned along the cycle.
We compute the partition functions of the orbifold
theories and observe that phase transitions occur
even in the zero 't Hooft coupling limit.
With the periodic boundary condition, the vacua of the
gauge theory are dual to various arrangements of 
$k$ NS5-branes.
With the anti-periodic boundary condition,
transitions between the vacua are dual to
localized tachyon condensations.
In particular, the mass of a deformed geometry
is compared with the Casimir energy for the dual
vacuum. We also obtain an index for the supersymmetric
orbifold theory.

\end{abstract}
\vfill
\vskip 0.5 truecm

\setcounter{footnote}{0}
\renewcommand{\thefootnote}{\arabic{footnote}}
\end{titlepage}

\newpage

\tableofcontents
\section{Introduction}
\label{Intoduction}

Recently the thermodynamics of large $N$ gauge theories
on compact spaces attract much attention.
On compact spaces the Gauss constraint restricts 
physical states into gauge invariant form,
and due to this fact the theories are in a
confinement phase at low temperature, and
undergo a deconfinement transition at a
critical temperature.
Moreover, the large $N$ gauge theories may
have their dual description in terms of string 
theory on an asymptotic Anti-de Sitter (AdS) space 
\cite{Maldacena}.
For example, the partition function for a large $N$
gauge theory on ${\mathbb R} \times S^3$ was
computed in \cite{Sundborg,AMMPR}, and it was shown that 
the partition function is of order ${\cal O}(1)$
at low temperature and of order ${\cal O}(N^2)$
above a critical temperature.
In the dual gravity theory, the phase transition
corresponds to the Hawking-Page transition 
\cite{HP,Wittent}, where the thermal AdS space is dominant at 
low temperature and the
AdS-Schwarzschild black hole is dominant at high temperature.

In this paper, we study the thermodynamics of ${\cal N}=4$ $U(N)$ 
super Yang-Mills theory on $\rs$.%
\footnote{The orbifold theory at zero temperature has been 
studied in \cite{LM}, see also \cite{ITT,ISTT}.} 
We construct the orbifold theory in the following way.
The manifold $S^3$ has a $U(1)$ symmetry along
a cycle, and the orbifold is constructed by 
dividing ${\mathbb Z}_k$ rotation along the cycle.
The background does not include any fixed point, but
has a non-trivial cycle due to the orbifold procedure.
We can introduce flux along the non-trivial cycle,
which gives non-trivial holonomies to the fields.
Therefore, the theory admits many vacua labelled by
the choice of flux, and this makes the phase diagram richer.
Along the non-trivial cycle, we can assign
periodic and anti-periodic boundary conditions to
fermions along the non-trivial cycle, and this leads to 
supersymmetric and non-supersymmetric theories at zero temperature.
We study the gauge theories
perturbatively with respect to the 't Hooft
coupling $\lambda = N g^2_{Ym}$.%
\footnote{
On the compact space, we have a tunable dimensionless
parameter $R \Lambda$, where $R$ is the radius of $S^3$ and
$\Lambda$ is a cut off scale.
If we take $R \Lambda \ll 1$, then the Yang-Mills coupling 
can be set small even at low energy.
Even with this fact we set $R=1$ for simplicity.}
In this paper, we consider the zero 't Hooft
coupling limit, where the theories reduce to
free field theories. Even in this limit we
observe phase transitions due to the compactness
of the base manifold.

In this orbifold case the dual gravity description is also available,
thus the phase diagram
can be extended into the strong coupling region.
For the case with the periodic boundary condition,
the dual geometry is the orbifold of the thermal AdS space
or its deformation by localized massless states in the low 
temperature phase. 
In the high temperature phase, the dual geometry is the orbifold
of the AdS-Schwarzschild black hole or its deformation.
If we perform the T-duality along the non-trivial
cycle, then we obtain $k$ NS5-brane configuration,
which is parametrized by the positions of $k$ NS5-branes \cite{LM}. 
For the case with the anti-periodic boundary condition, 
there are localized tachyons 
at the fixed point of the thermal AdS orbifold in the
low temperature phase.
The condensation of localized tachyon may resolve the
orbifold singularity like \cite{APS} and lead to
the deformed geometry obtained in \cite{CM1,CM2}
called as Eguchi-Hanson soliton.
In other words, the gauge theory gives the dual
picture of the localized tachyon condensation
discussed in \cite{APS}.\footnote{Previous attempts to
apply the AdS/CFT correspondence to the localized tachyon
condensation have been given in \cite{TZ,CST,AS,DKR1,DKR2}.}
At enough high temperature, the dual geometry should
be the orbifold of the AdS-Schwarzschild black hole,
and there are no localized tachyons in the geometry.

Following the analysis in \cite{Sundborg,AMMPR},
we obtain the partition functions for the gauge theories
in terms of a matrix integral. 
At low temperature, we find the leading contribution 
comes from the Casimir energy of the theories on 
$S^3/{\mathbb Z}_k$. For the case with the periodic boundary
condition, the Casimir energy is the same for all the vacua 
and the same as the mass of the thermal AdS orbifold.
For the case with the anti-periodic boundary condition, 
the Casimir energy is smallest for the vacuum dual 
to the deformed geometry. Interestingly, the Casimir
energy is roughly $4/3$ times the mass of the 
Eguchi-Hanson soliton.
In the high temperature limit, the partition function behaves
in the same way for all possible holonomies and spin
structures. Near the critical temperature, we can
perform an analytic computation by using the Gross-Witten
ansatz \cite{GW} as an approximation, and we can discuss the
dominant contribution to the total partition function.

The organization of this paper is as follows.
In the next section, we define the orbifold gauge 
theories on $\rs$ with generic holonomy by
utilizing the standard orbifold method as in \cite{DM}.
We compute the partition function for the orbifold gauge
theories by following \cite{Sundborg,AMMPR}.
In section \ref{phase} we analyze the partition function
and discuss the phase structure.
We observe that the dominant contribution comes from the 
Casimir energy at low temperature, 
and we compute the Casimir energy for the case 
with general holonomy.
Near the critical temperature we solve the partition function
analytically by making use of the analysis in \cite{GW}.
At very high temperature, we obtain the partition function
as an expansion of the temperature $T$, which does not
depend on the choice of holonomy.
In section \ref{gravity} we analyze the large 't Hooft coupling
limit in the dual gravity description.
We discuss the relations to the arrangement of $k$ NS5-branes for
the case with the periodic boundary condition 
and to the localized tachyon condensation for the case
with the anti-periodic boundary condition.
Section \ref{conclusion} is devoted to conclusion and discussions.
In appendix A we compute the partition function of
single scalar particle as an example. In appendix B the index 
proposed in \cite{KMMR} is computed for the supersymmetric orbifold
theory.%
\footnote{Similar computations were done in \cite{Nakayama1}
for quiver gauge theories, which are constructed as orbifolds
different from ours.}


\section{Orbifold gauge theories}
\label{orbifold}

We consider ${\cal N} = 4$ super Yang-Mills theories on $\rs$ 
with large $N$ $U(N)$ gauge symmetry. 
The 3-sphere $S^3$ has $SO(4) \simeq SU(2)_1 \times SU(2)_2 $ isometry, 
and we divide the gauge theories by $4 \pi /k$ rotation along
the $\chi$-cycle of $U(1)_{\chi} \subset SU(2)_2$.
Since $\pi_1 (S^3/{\mathbb Z}_k) = {\mathbb Z}_k$,
we can assign a non-trivial holonomy matrix 
$V = P \exp i \oint A_{\chi}$ along the non-trivial cycle.
Using the $U(N)$ gauge transformation, we can set $V$ as a diagonal
matrix $V = {\rm diag} (\Omega_1, \cdots , \Omega_N)$.
Because of the condition $V^k = 1$, the element should be
a $k$-th root of unity $\Omega^k_j = 1$.
Therefore, we can label the vacua by  
$(n_0, \cdots , n_{k-1})$ with $N = \sum_{I=0}^{k-1} n_I$, where 
$n_I$ represents the number of $j$ such that $\Omega_j = \omega^I$ 
$(\omega = \exp (2 \pi i/k))$.
Then the orbifold theories are defined by projecting
the Hilbert space into the orbifold invariant subspace as below.
In the following we consider two special vacua. One is
the ${\mathbb Z}_k$ symmetric vacuum%
\footnote{In this case $N$ is assumed to be $N = k {\mathbb Z}$. 
However, for large $N$ and finite $k$, the difference from the 
general $N$ case should be negligible.}  with $n_I = N/k$ for all $I$.
Due to the ${\mathbb Z}_k$ symmetry the dual geometry can be
identified as the standard orbifold.
The other vacua do not preserve the ${\mathbb Z}_k$ symmetry,
thus the dual geometry should be a deformation of the orbifold.
The other important vacuum is with the trivial holonomy $V=1$ or
equivalently $n_0 = N$. In this case the ${\mathbb Z}_k$ symmetry
is maximally broken.

\subsection{Partition function}

We would like to compute the partition function of
gauge invariant operator in the orbifold theories.
In general, the counting of gauge invariant operator
is an involved task. Fortunately, it was shown in 
\cite{Sundborg,AMMPR} 
that the partition function of gauge invariant
operator can be written in terms of single-particle 
partition functions as
\begin{align}
 Z(x) = \int [ dU] \exp \left[ 
  \sum_{\cal R} \sum_{n=1}^{\infty} \frac{1}{n}
      z^{\cal R} (x^n) \chi_{\cal R} (U^n)  \right] ~.
      \label{pf1}
\end{align}
At this stage, a gauge group $G$ could be arbitrary,
and the sum is taken over all representation ${\cal R}$
of the gauge group $G$. We denote $\chi_{\cal R} (U)$ as the
character for representation ${\cal R}$, and $[dU]$ as the Haar
measure for the group element $U$. 
The partition function of single particle
in the representation ${\cal R}$ is computed by
\begin{align}
 z^{\cal R} (x) = \sum_E x^E ~,
 \label{sp}
\end{align}
where $E$ denotes the energy eigenvalue.
The condition of gauge invariance comes from the integral
over $U$, and the variable $U$ will be identified as
the holonomy matrix along the thermal cycle.

We can adopt any gauge group $G$ and representation ${\cal R}$
in the formula \eqref{pf1}, and several interesting
examples have their dual gravity description.
The most famous one arises from the $N$ D3-brane
worldvolume theory, which is dual to superstrings on
$AdS_5 \times S^5$.
The theory is ${\cal N}=4$ super Yang-Mills theory on 
${\mathbb R} \times S^3$ with gauge group $G=U(N)$, where
the states are in the adjoint representation.
Our interest is on the orbifold gauge theories
with holonomy along the non-trivial cycle, where
the existence of the holonomy
$(n_0, \cdots , n_{k-1})$ breaks the gauge symmetry
into $G = \prod_{I=0}^{k-1} U(n_I)$.%
\footnote{The diagonal $U(1)$ parts of each $U(n_I)$ may 
be decoupled from the rest, but the difference can be ignored 
when $n_I$ are very large. See, however, appendix B for the case of 
an index.} 
With respect to the broken gauge group, the states are in the
adjoint representation for $U(n_I)$ or in the bi-fundamental
representation $( n_I , \bar n_J )$ for 
$U(n_I) \times U(n_J)$.

The spectrum of the orbifold theories can be obtained
by projecting the spectrum on $S^3$ into the orbifold
invariant subspace. The spectrum on $S^3$ can be obtained
from the spherical harmonic analysis as in \cite{Cutkosky}.
The theory includes 6 scalers, a gauge field 
and 4 Majorana fermions.
The scalars can be expanded by the scalar spherical 
harmonics $S_{j,m,\bar m}(\Omega)$, where $\Omega$
represents the coordinates of $S^3$.
The eigenfunctions of Laplace operator on $S^3$ are given as
\begin{align}
 \nabla^2 S_{j,m,\bar m} (\Omega) = - j(j+2) S_{j,m,\bar m} (\Omega)~.
\end{align}
The labels $(m,\bar m)$ are eigenvalues of $J^3$ and $\bar J^3$
for $SU(2)_1$ and $SU(2)_2$, and they
run $ - j/2, - j/2 +1 , \cdots , j/2 -1 ,j/2$.
The projection into the orbifold invariant modes is
performed by the projection operator
$P = \frac{1}{k} \sum_{I=0}^{k-1} \Gamma^I$,
where $\Gamma$ represents the orbifold action.
For a bi-fundamental state $(n_I,\bar n_J)$ or an adjoint state with $I = J$,
the orbifold action is given by
\begin{align}
 \Gamma = e^{4 \pi i \bar J_3/k} \omega^{I - J} ~,
 \label{orbifolda}
\end{align}
where $\omega = \exp (2 \pi i / k)$.
The orbifold action consists of two parts.
The first part is the phase shift due to the ${\mathbb Z}_k$ 
rotation along the $\chi$-cycle.
The second part is the holonomy for the bi-fundamental
state $(n_I , \bar n_J)$.
We can see from \eqref{orbifolda} that
the orbifold invariant modes are restricted
to $ 2 \bar m = J - I$ mod $k$. As a notation
we define $0 \leq L < k$ subject to $L = J - I$ mod $k$.
Now we can compute the partition function for the
single scalar particle \eqref{sp} as
\begin{align}
\begin{aligned}
 z^{I,J}_S (x) &= \frac{(x^{L+1} - x^{L+3} + x^{- L +1} - x^{- L + 3}) k x^{k} }
                {(1 - x^2)^2 ( 1 - x^k)^2} \\
          &+ \frac{ (L+1) x^{L+1} - (L-1) x^{L+3} 
                 - (L-1) x^{k - L +1} +(L+1) x^{k - L + 3} }
                {(1 - x^2)^2 ( 1 - x^k)} ~.
                \label{zls}
\end{aligned}
\end{align}
See appendix A for the detail.
We have used the fact that the energy is given by $E=j+1$
for a scaler on $S^3$ conformally coupled to gravity.

We move to the gauge field, which is expanded by the vector spherical
harmonics $V^{\pm}_{j,m,\bar m} (\Omega)$.%
\footnote{A longitudinal mode is expanded by the scalar 
spherical harmonics as $\vec \nabla S$, and we do not consider it.} 
We use the notation such that the vector index is
contracted with an auxiliary unit vector $\hat \xi_{\mu}$ as
$V^{\pm}_{j,m,\bar m;\mu} \hat \xi^{\mu}$.
The vector spherical harmonics $V^{+}_{j,m,\bar m}$ and
$V^{-}_{j,m,\bar m}$ belong to the representations 
$(j_1 , j_2 )= (\frac{j + 1}{2},\frac{j - 1}{2})$ and
$(\frac{j - 1}{2},\frac{j + 1}{2})$, respectively.
The eigenvalues of Laplace operator on $S^3$ are
\begin{align}
 \nabla^2 V^{\pm}_{j,m,\bar m} (\Omega) = - (j+1)^2 V^{\pm}_{j,m,\bar m} (\Omega) ~.
\end{align}
The orbifold action to the vector spherical harmonics is the
same as in the scalar case \eqref{orbifolda}, since
the vector index is contracted with an auxiliary unit vector.
Therefore, the orbifold projection allows only the modes with
$ 2 \bar m = J - I$ mod $k$
for the bi-fundamental state with $(n_I , \bar n_J)$.
The partition function is then given by
\begin{align}
\begin{aligned}
 z^{I,J}_{V^+} (x) &= \frac{(x^{L+2} - x^{L+4} + x^{- L +2} - x^{- L + 4}) k x^{k} }
                {(1 - x^2)^2 ( 1 - x^k)^2} \\
          &+ \frac{ (L+3) x^{L+2} - (L+ 1) x^{L+4} 
                 - (L-3) x^{k - L +2} +(L-1) x^{k - L + 4} }
                {(1 - x^2)^2 ( 1 - x^k)}
\end{aligned}
\end{align}
for $V^{+}_{j,m,\bar m}$ and
\begin{align}
\begin{aligned}
 z^{I,J}_{V^-} (x) &= \frac{(x^{L} - x^{L+2} + x^{- L} - x^{- L + 2}) k x^{k} }
                {(1 - x^2)^2 ( 1 - x^k)^2} \\
          &+ \frac{ (L-1) x^{L} - (L - 3) x^{L+2} 
                 - (L+1) x^{k - L} +(L+3) x^{k - L + 2} }
                {(1 - x^2)^2 ( 1 - x^k)}
\end{aligned}
\end{align}
for $V^{-}_{j,m,\bar m}$. We have defined $L = J - I$ mod 
$k$ ($0 \leq L < k$) as before.

Fermions are expanded by the spherical harmonics 
$F^{+}_{j,m,\bar m} (\Omega)$  and  $F^{-}_{j,m,\bar m} (\Omega)$, which
belong to $(j_1,j_2) = (\frac{j}{2},\frac{j-1}{2})$
and $(\frac{j-1}{2},\frac{j}{2})$.
The spinor index is again contracted with an auxiliary
spinor $\chi^{\alpha}$ as 
$F^{\pm}_{j,m,\bar m ; \alpha} \chi^{\alpha}$.
The eigenvalues of Laplace operator on $S^3$ are
\begin{align}
 \nabla^2 F^{\pm}_{j,m,\bar m} (\Omega)= - (j+ {\textstyle \frac12 })^2 F^{\pm}_{j,m,\bar m} (\Omega)~.
\end{align}
For fermions we can assign two types of boundary conditions
along the non-trivial cycle.
The orbifold action depends on the boundary condition as
\begin{align}
 \Gamma = \pm e^{4 \pi i \bar J_3/k} \omega^{I - J} ~,
\end{align}
where $+$ and $-$ means the periodic and the anti-periodic
boundary conditions, respectively.
For the periodic boundary condition, the orbifold invariant 
modes are given by those with $2 \bar m = J - I$ mod $k$.
The anti-periodic boundary condition can be assigned only 
for even $k$, and the restriction is shifted by $k/2$ as
$2 \bar m = J - I + k/2$ mod $k$.
For the periodic boundary condition,
the partition function can be computed as
\begin{align}
\begin{aligned}
 z^{I,J}_{F^+} (x) &= \frac{ (x^{L+\frac32} - x^{L+\frac72} 
                + x^{- L +\frac32} - x^{- L + \frac72}) k x^{k} }
                {(1 - x^2)^2 ( 1 - x^k)^2} \\
          &+ \frac{ (L+2) x^{L+\frac32} - L x^{L+\frac72} 
                 - (L-2) x^{k - L + \frac32} + L x^{k - L + \frac72}}
                {(1 - x^2)^2 ( 1 - x^k)}
\end{aligned}
\end{align}
for $F^+_{j,m,\bar m}$ and
\begin{align}
\begin{aligned}
 z^{I,J}_{F^-} (x) &= \frac{ (x^{L+\frac12} - x^{L+\frac52} 
                + x^{- L +\frac12} - x^{- L + \frac52}) k x^{k} }
                {(1 - x^2)^2 ( 1 - x^k)^2} \\
          &+ \frac{ L x^{L+\frac12} - (L-2) x^{L+\frac52} 
                 - L x^{k - L + \frac12} + (L+2) x^{k - L + \frac52} }
                {(1 - x^2)^2 ( 1 - x^k)}
\end{aligned}
\end{align}
for $F^-_{j,m,\bar m}$.
For the anti-periodic boundary condition, we should use
$L = J - I + k/2$ mod $k$ with $0 \leq L < k$
instead of $L = J - I$ mod $k$.

Now we can write up explicitly the partition function \eqref{pf1}
for ${\cal N}=4$ super Yang-Mills theories on $\rs$
with holonomy $(n_0 , \cdots , n_{k-1})$.
The total partition function is given by summing over all the vacua.
It is useful to use the formula for the character of bi-fundamental 
representation as 
$\chi_{( n_I ,  \bar n_J)}(U) = \Tr U_I \Tr U^{\dagger}_J$,
where the trace is taken in the fundamental representation.
The partition function is then given by
\begin{align}
 Z(x) = \int [ \prod_{I} dU_I ] \exp \left[ 
  \sum_{I,J} \sum_{n=1}^{\infty} \frac{1}{n}
      z^{I,J}_n (x)  \Tr (U_I^n) \Tr (U_J^{\dagger n}) \right] ~,
      \label{pf2}
\end{align}
where the single-particle partition function is summarized as
\begin{align}
       z^{I,J}_n (x) = 6 z_S^{I,J} (x^n)  + z_{V^+}^{I,J} (x^n)  
            + z_{V^-}^{I,J} (x^n) + (-1)^{n+1} 4 [z_{F^+}^{I,J} (x^n)
             + z_{F^-}^{I,J} (x^n) ]
\label{zsuper}
\end{align}
for the case with the periodic boundary condition and
\begin{align}
       z^{I,J}_n (x) = 6 z_S^{I,J} (x^n)  + z_{V^+}^{I,J} (x^n)  
            + z_{V^-}^{I,J} (x^n) 
     + (-1)^{n+1} 4 [z_{F^+}^{I,J + \frac{k}{2}} (x^n)
     + z_{F^-}^{I,J + \frac{k}{2}} (x^n) ]
\label{znon}
\end{align}
for the case with the anti-periodic boundary condition.

Finally let us remark on the difference from the
D-brane worldvolume theories localized at the fixed point of 
${\mathbb C}^n/\Gamma$ with $n=2,3$ \cite{DM}.
Since the orbifold action acts trivially to the worldvolume
in those cases, only bi-fundamental matters with $I,I \pm 1$ and 
adjoint gauge fields (and matters) are left under the orbifold projection.
On the other hand, the orbifold action rotates $S^3$ by
$4\pi/k$ in our case, there is a $\bar m$ dependent phase
in \eqref{orbifolda}.
Due to this effect,  bi-fundamental states with every 
pairs of $I,J$ (and adjoint states with $I=J$) 
survive the projection each for matters, 
gauge field and fermions.
The difference would be significant if we compare our case 
with the duality between superstrings on $AdS_5 \times S^5/\Gamma$ 
and the gauge theory coming from 
D3-branes at the fixed point of orbifold 
action $\Gamma$ \cite{KS}.

\subsection{Path integral formulation}

In the previous subsection, we have obtained the partition function
of gauge invariant operator \eqref{pf2} in terms of
integral over the group manifolds. However, we cannot
determine the overall pre-factor in the formulation.
In this subsection, we re-derive the partition function
in the path integral formulation.
In this derivation, we obtain the normalization
depending on the Casimir energy of the gauge theories
on $\rs$. The Casimir energy will be important when we
consider the phase structure at low temperature.
Moreover, we can identify $U_I$ as the holonomy
matrix for $U(n_I)$ gauge group along the thermal cycle.

The path integral for the partition function with a finite temperature
$T$ may be computed on $S^1 \times S^3/{\mathbb Z}_k$, where $S^1$ 
is the thermal cycle with periodicity $\beta = 1/T$.
Along the thermal cycle we assign the anti-periodic boundary
condition for the fermions.%
\footnote{Due to this boundary condition, supersymmetry is
always broken at a finite temperature even for the theory 
supersymmetric at zero temperature. 
}
We start from fixing the gauge symmetry and then introduce the 
Faddeev-Popov determinant conjugate to the gauge fixing.
We adopt the Coulomb gauge
\begin{align}
 \nabla_a A^a = 0 
 \label{gauge1}
\end{align}
with $\nabla_a$ as covariant derivatives along the 
$S^3$ direction $(a=1,2,3)$. 
If we do not include a non-trivial holonomy,
then there are spatially constant modes of the gauge field.
The presence of holonomy $(n_0 , \cdots , n_{k-1})$
breaks the gauge group into $\prod_I U(n_I)$ and
spatially constant modes are left only for $\prod_I U(n_I)$.%
\footnote{The projection under the orbifold action \eqref{orbifolda}
removes spatially constant modes with $\bar m = 0$ for $I \neq J$ 
sectors.}
The time-dependence of these modes is not fixed by the Coulomb 
gauge \eqref{gauge1}, and we fix these degrees by
\begin{align}
 \partial_t \alpha^I &= 0 ~,
  &\alpha^I &= \frac{1}{{\rm Vol}(\sz)} \int d \Omega A^I_t ~,
  \label{gauge2}
\end{align}
where the integration is performed over $\sz$.

First we consider the Faddeev-Popov determinant conjugate to
\eqref{gauge2}, which is given by
\begin{align}
 \Delta^I_{\rm FP} &= \det {}' ( \partial_t D^I_t )~,
 & D^I_t  & = \partial_t  - i [ \alpha^I , ~~ ] ~.
\end{align}
The determinant is taken over the non-zero modes.
Diagonalizing the zero modes as
$\alpha^I = {\rm diag} ( \alpha_1^I, \cdots \alpha_{n_I}^I)$,
the measure can be written as
\begin{align}
 d \alpha^I =  \prod_i d \alpha_i^I \prod_{i  , j} | \alpha^I_i - \alpha^I_j |
 ~,
\end{align}
where the Van der Monde determinant arises from
the integration over the off diagonal elements.
Now that the bosonic modes are periodic along the thermal cycle,
they can be expanded by the function 
$\exp (2 \pi i n t / \beta )$ with $n \in {\mathbb Z}$.%
\footnote{For fermionic modes, we should replace $n$ by $n + 1/2$
since we have assigned anti-periodic boundary condition along the thermal
cycle.}
Thus the determinant can be written as
\begin{align}
 \Delta^I_{\rm FP} = \prod_{i,j} \prod_{n \neq 0} \frac{2 \pi i n }{ \beta }
 \left( \frac{2 \pi i n }{ \beta }- i ( \alpha^I_i - \alpha^I_j) \right) ~.
\end{align}
With the help of the formula 
$\prod_{n=1}^{\infty} (1 - x^2/n^2) = \sin \pi x / (\pi x )$,
we find up to an overall factor
\begin{align}
 [ d U_I ]  =  d \alpha^I \Delta^I_{\rm FP} 
         =  \prod_i d \alpha^I_i 
 \prod_{i < j} \sin ^2 \left( \frac{\beta (\alpha^I_i - \alpha^I_j )}{2} \right) ~,
\end{align}
which is the Haar measure of 
$U_I = \exp (i \beta \alpha^I )$.

The Faddeev-Popov determinant conjugate to \eqref{gauge1} is given by
\begin{align}
 \det \nabla_a D^a = \int [d c d \bar c] 
\exp ( - \bar c \nabla_a D^a c ) ~,
\end{align}
which should be added to the action of ${\cal N}=4$ super
Yang-Mills theory.
Notice that the ghosts are expanded by the scalar spherical 
harmonics projected into the orbifold invariant modes.
After integrating over the massive modes including
the $c$-ghosts, the partition function is given in 
terms of integral over the zero modes as
\begin{align}
 Z(T) = \int [ \prod_I d U_I ] e^{ - S ( U ) } ~.
 \label{pathpf}
\end{align}
Let us first compute the contribution from the gauge field
and the $c$-ghosts.
Since the longitudinal modes $\vec \nabla S$ and $A_t$ (except the
zero modes $\alpha^I$) are expanded by the scalar 
spherical harmonics,  the contributions to the path 
integral from the $c$-ghosts and the longitudinal modes 
cancel out. Therefore, the contribution
reduces to the Gaussian integral over the vector spherical harmonics, 
which is evaluated as
\begin{align}
 S ( U ) = \frac12 \sum_{I,J} 
  \sum_E [ n_{V^+}^{I,J} (E) + n_{V^-}^{I,J} (E) ] \ln \det ( - D_t^2 + E^2 ) ~.
\end{align}
We denote $n_{V^{\pm}}^{I,J}(E)$ as the degeneracy of eigenstates with $E$
in the representation $(n_I , \bar n_J)$.
Following the computation in \cite{AMMPR}, we find
\begin{align}
\begin{aligned}
 S (U) &= \frac12 \sum_{I,J} \beta n_I n_J \sum_E 
     [ n_{V^+}^{I,J} (E) + n_{V^-}^{I,J} (E) ] E \\
     &- 
     \sum_{I,J} \sum_{n=1}^{\infty} \frac{1}{n} 
    [ z^{I,J}_{V^+} ( e^{-n/T}) +   z^{I,J}_{V^-} ( e^{-n/T}) ]
     \Tr (U_I^n) \Tr (U_J^{\dagger n}) ~.
\end{aligned}
\end{align}
In the same way, we can compute the contributions from
scalars and fermions and summarize all the contributions as%
\footnote{In the expression of \eqref{pathpf} 
with \eqref{eff}, the normalization has been set by dividing
holonomy independent factors.
Introducing the holonomy $(n_0 , \cdots , n_I)$
along the $\chi$-cycle, the lowest modes 
in $U(N)/\prod_I U(n_I)$ become space-dependent, and hence 
they can be
 fixed by the Coulomb gauge \eqref{gauge1} instead of 
\eqref{gauge2}.
The normalization of each Faddeev-Popov determinants
may depend on the choice of holonomy along the $\chi$-cycle,  
but the sum of both should not. }
\begin{align}
 S (U) &= \beta V_0 -
     \sum_{I,J} \sum_{n=1}^{\infty} \frac{1}{n} z^{I,J}_n (e^{-1/T})
     \Tr (U_I^n) \Tr (U_J^{\dagger n}) ~,
     \label{eff}
\end{align}
where $z_n^{I,J}$ are the single-particle partition functions
\eqref{zsuper} or \eqref{znon}.
The first term is the Casimir energy
\begin{align}
 V_0 = \frac12 \sum_{I,J} n_I n_J \sum_E 
     [ 6 n_S^{I,J} (E) + n_{V^+}^{I,J} (E) + n_{V^-}^{I,J} (E) 
       - 4  n_{F^+}^{I,J} (E) - 4 n_{F^-}^{I,J} (E)  ] E ~.
       \label{cas}
\end{align}
Compared with the expression of \eqref{pf2},
the integral variables $U_I$ are identified with
the holonomy matrices $U_I = \exp(i \beta \alpha^I)$
with respect to the gauge group $\prod_I U(n_I)$.
In this way we can see that the previous expression
only includes the finite temperature contribution.
The zero temperature contribution, which comes from the
Casimir energy, should be included in the partition function.


\section{Phase transitions of the gauge theories}
\label{phase}

In the previous section, we have obtained the partition function
of gauge invariant operator in terms of integral over $U_I$
as \eqref{pathpf} with \eqref{eff}. 
In this section, we perform the $U_I$ integral in the
large $N$ limit%
\footnote{Finite $N$ effects may be examined by following the 
analysis in \cite{Liu,ALW,ABMW}. }
and examine the phase structure of the orbifold theories. 
For large $N$ and fixed $k$, it is natural to assume that $n_I$ 
in the label of holonomy $(n_0, \cdots , n_{k-1})$ are very large.
In case that some of $n_I$ are very small, then
they may be set zero in this limit.
We consider two specific vacua in the following.
One is the ${\mathbb Z}_k$ symmetric holonomy vacuum
with $n_I = N/k$ and the other is the trivial holonomy vacuum
with $n_0 = N$. In these cases our assumption is valid.
In the next subsection we study the low temperature
phase, and in subsection \ref{casimir} we focus on the Casimir
energy contribution. In subsection \ref{critical} we obtain
an analytic expression under an approximation near 
the critical temperature.
In subsection \ref{high} we take the high temperature limit, 
where the analysis becomes simpler.

\subsection{Critical temperatures}
\label{low}

It is convenient to diagonalize the eigenvalues of holonomy 
matrix $U_I$ as $\exp ( i \theta_{I, i})$ with 
$- \pi \leq \theta_{I,i} < \pi$.%
\footnote{The eigenvalue $\theta_{I,i}$ is related with
the zero modes in \eqref{gauge2} as
$\theta_{I,i} = \beta \alpha^I_i$.}
For large $n_I$ the discrete elements may be replaced by
a continuous parameter $\theta_I$ with a density 
$\rho^I (\theta_I)$.
The density has to satisfy $\rho^I (\theta_I) \geq 0$ and
the normalization is set as 
$\int_{- \pi}^{\pi} \rho^I (\theta_I) d \theta_I = 1$.
In this approximation the effective action \eqref{eff} 
becomes
\begin{align}
\begin{aligned}
 S [\rho^I (\theta_I) ] = \beta V_0 - \sum_{I, J } n_I n_J 
 \int d\theta_I d \theta'_J
 \rho^I (\theta_I ) \rho^J (\theta '_J ) \Bigl[ \delta_{I,J}
 \ln \left| \sin \left( \frac{\theta_I - \theta '_J}{2} \right) \right|
 \\ + \sum_{n=1}^{\infty} \frac{1}{n} z^{I,J}_n (x) 
   \cos ( n ( \theta_I - \theta '_J)) \Bigr]
   \label{eff2}
\end{aligned}
\end{align}
with $x = e^{- 1/T}$.
The first term in the bracket arises from the change of measure
$[d U_I] \to [ d \theta_{I,i} ]$.
In terms of the Fourier transform 
$\rho^I_n = \int d \theta_I \rho^I (\theta_I) \cos ( n \theta_I )$,%
\footnote{We assume that $\theta_I$ is distributed symmetrically around 
$\theta_I = 0$.}
the effective action \eqref{eff2} is given by
\begin{align}
 S[ \rho^I_n ]
  = \beta V_0 + \sum_{I,J} n_I n_J 
    \sum_{n=1}^{\infty}  \rho^I_n \bar \rho^J_n V^{I,J}_n (x) ~,
    \label{St}
\end{align}
where
\begin{align}
 V^{I,J}_n (x) = \frac{1}{n} ( \delta_{I,J} - z^{I,J}_n(x) ) ~.
    \label{V^L}
\end{align}

At enough low temperature, 
the repulsive force coming from the first term of \eqref{V^L}
dominates, and the uniform distribution 
$\rho_n^I =0$ for $n \geq 1$ is the classical solution to 
the effective action \eqref{St}.
Therefore, there are no order ${\cal O}(N^2)$ (nor order ${\cal O}(N)$) 
contributions from the effective action except for the Casimir energy term $\beta V_0$.
An order ${\cal O}(1)$ contribution comes from the Gaussian
integral as
\begin{align}
 \prod_{n = 1}^{\infty} \frac{1}{ \det \left(n_I n_J V^{I,J}_n (x) \right)} ~,
 \label{gauss}
\end{align}
where the determinant is over the labels $0 \leq I,J < k$.
As the temperature increased, the second term of \eqref{V^L}
contributes to the potential, and the determinant would vanish
at a critical temperature $x_c = \exp ( - 1/T_c )$.
Above the critical temperature, the distribution becomes non-uniform
and the classical contribution is of order ${\cal O}(N^2)$.

Let us examine two concrete examples.
We start from the trivial holonomy case,
where the action \eqref{St} reads
\begin{align}
 S[ \rho^0_n ]
  = \beta V_0 + N^2 
    \sum_{n=1}^{\infty}  | \rho^0_n |^2
  \frac{1}{n} ( 1 - z^{0,0}_n (x) ) ~.
\end{align}
At enough low temperature, the coefficients of $|\rho^0_n|^2$
are positive, and $\rho^0_n = 0$ for $n \geq 1$ is the saddle point.
Now that the coefficients are $1 \times 1$ matrices,
the determinant \eqref{gauss} is simply
\begin{align}
 \prod_{n = 1}^{\infty} \frac{1}{ 1 - z^{0,0}_n (x) } ~.
\end{align}
We have changed the normalization such that only the Casimir energy
term is left at $x=0$.
As the temperature increased, the coefficients of $|\rho^0_n|^2$ 
become smaller, and at a critical temperature $T = T_c$, a
coefficient vanishes. Since the single-particle
partition function is a monotonically increasing function
of $x$, the first zero comes from the $n=1$ part when
$1 - z^{0,0}_1 (x) = 0$.
The critical temperatures  $x_c$ and $T_c$ are summarized 
for small $k$ in Table \ref{table}.
\begin{table}
\centerline{
 \begin{tabular}{|c||c|c||c|c|}
  \hline
  $k$ & $x_c$ (periodic) &  $T_c$ (periodic) & 
  $x_c$ (anti-periodic) &  $T_c$ (anti-periodic) \\
  \hline 
   2 & 0.095663 & 0.426090 & 0.095663 & 0.426090 \\
   4 & 0.104448 & 0.442661 & 0.127999 & 0.486445 \\
   6 & 0.104684 & 0.443104 & 0.139545 & 0.507777 \\
   8 & 0.104689 & 0.443113 & 0.142528 & 0.513290 \\
   10 & 0.104689 & 0.443113 & 0.143136 & 0.514414 \\ 
  \hline
 \end{tabular}}
 \caption{\it The critical temperatures $x_c = \exp(-1/T_c)$ and
 $T_c$ for the ${\cal N}=4$ super Yang-Mills theories on 
${\mathbb R} \times S^3/{\mathbb Z}_k$ with the trivial holonomy $V=1$.
We set $k=2,4,6,8,10$. The periodic and 
anti-periodic boundary conditions are assigned for the fermions 
along the $\chi$-cycle.}
 \label{table}
\end{table}
We should note that for $k=1$ the critical temperature
reduces to the one for ${\mathbb R} \times S^3$ case \cite{Sundborg,AMMPR} as 
$x_c = 7 -4 \sqrt{3} = 0.071797$ or $T_c = 0.379663$.

Another interesting case may be with the ${\mathbb Z}_k$ 
symmetric holonomy $n_I = N/k$ for all $I$.
In this case the action \eqref{St} is given by
\begin{align}
 S[ \rho^I_n ]
  = \beta V_0 + \frac{N^2}{k^2} \sum_{I,J}  
    \sum_{n=1}^{\infty}  \rho^I_n \bar \rho^J_n \frac{1}{n}
     ( \delta_{I,J} - z^{I,J}_n(x) ) ~.
\end{align}
Notice that the coefficients of $\rho^I_n$ take the form of a circulant
matrix since $z^{I,J}_n$ only depends on the
difference $I - J$.
Using the formula for a circulant determinant \eqref{cd},
the determinant \eqref{gauss} can be written in a compact 
form as
\begin{align}
 \prod_{n=1}^{\infty} \prod_{I = 0}^{k - 1} \frac{1}
  { 1 - \sum_{J=0}^{k-1} \omega^{I J} z^{0,J}_n (x) } ~.
\end{align}
As before it is enough to focus on the $n = 1$ factors.
If we increase the temperature, then the denominator
diverges when $\sum_J z^{0,J}_1 (x) = 1$.
Among the other factors, this factor gives the divergence
with smallest $x_c$ since $z^{0,J}_1 (x)$ is positive for all $J$.
Remarkably the critical temperature is the same
for all $k$ and for both the spin structures as 
$x_c = 0.071797$ or $T_c = 0.379663$ as in the
${\mathbb R} \times S^3$ case.
Actually this is an expected result since the sum of all 
sectors with weight one $\sum_J z^{0,J}_1 (x)$ is the same as 
the single-particle partition function for the ${\mathbb R} \times S^3$ case. 

\subsection{Casimir energies}
\label{casimir}

At low temperature, the determinant \eqref{gauss} is of order
${\cal O}(1)$ and the contribution from the Casimir energy 
\eqref{cas} is dominant.
The Casimir energy is an important quantity 
since it is supposed to correspond to the mass of 
the dual geometry.
In order to compute Casimir energy, we have to sum over
infinitely many states, and this may lead to a divergent result.
Thus we have to choose a regularization, but it is a subtle
problem for quantum field theory on a curve background \cite{BD}.
Fortunately, we will find that the Casimir energies in our case
are finite, thus we do not need to worry about this difficult issue.

In order to regularize the infinite sum in the Casimir energy
\eqref{cas}, we first introduce a cut off factor $e^{- E/\mu}$ as
\begin{align}
\sum_E ( 6 n^{I,J}_{S} + n^{I,J}_{V^+}
 + n^{I,J}_{V^-} - 4 n^{I,J}_{F^+} - 4 n^{I,J}_{F^-} )
  E e^{- E / \mu} ~,
\end{align}
and finally take the limit $\epsilon = 1/\mu \to 0$. 
This regularization may be justified by the fact that
no divergent terms are left in the final results.
The above regularization can be realized by using the
identity $(A=S,V^{\pm},F^{\pm})$
\begin{align}
\sum_E n^{I,J}_{A} (E) E e^{- \epsilon E}
 = - \frac{d}{d \epsilon} z^{I,J}_{A} (e^{- \epsilon}) ~,
 \label{reg}
\end{align}
so we need the expansion of single-particle partition functions by 
$\epsilon$ up to ${\cal O}(\epsilon)$ as
\begin{align}
 z_S^{I,J} (e^{- \epsilon}) &\sim
 \frac{2}{\epsilon ^3 k} -  \frac{7 \epsilon}{180 k}
 + \frac{\epsilon k}{36} + \frac{\epsilon k^3}{360}
 - \frac{\epsilon L (k - L)}{6 k}
 - \frac{\epsilon L^2 (k - L)^2}{12 k} ~, 
\\
 z_{V^{\pm}}^{I,J} (e^{- \epsilon}) &\sim
 \frac{2}{\epsilon ^3 k} - \frac{1}{\epsilon k} 
 \pm \frac{k}{6} \pm \frac{1}{3 k} \mp L \pm \frac{L^2}{k} 
 + \left( \frac{1 \mp 1}{2} \right) \delta_{L,0} 
 + \frac { 2 \epsilon}{45 k}
 - \frac{ 5 \epsilon k}{36} + \frac{\epsilon k^3}{360} \nonumber \\
 &+ \frac{5 \epsilon L (k - L )}{6 k}
 - \frac{\epsilon L^2 ( k - L)^2}{12 k} ~, 
\\
 z_{F^\pm}^{I,J} (e^{- \epsilon}) &\sim
 \frac{2}{\epsilon ^3 k} - \frac{1}{4 \epsilon k} 
 \pm \frac{k}{12} \mp \frac{1}{12 k} \mp \frac{L}{2} \pm \frac{L^2}{2 k} 
 + \frac { 83 \epsilon}{2880 k}
 - \frac{ \epsilon k}{72} + \frac{\epsilon k^3}{360} \nonumber \\
 &+ \frac{\epsilon L ( k - L)}{12 k} 
 - \frac{\epsilon L^2 ( k - L)^2}{12 k}  ~.
\end{align}
Below we discuss the cases with the periodic and the anti-periodic
boundary conditions separately.

For the case with the periodic boundary condition, we have
\begin{align}
\sum_E ( 6 n^{I,J}_{S} + n^{I,J}_{V^+}
 + n^{I,J}_{V^-} - 4 n^{I,J}_{F^+} - 4 n^{I,J}_{F^-} )
  E e^{- \epsilon E} =  \frac{3}{8 k} + {\cal O}(\epsilon) ~.
\end{align}
As mentioned before, the final expression does not depend on
the cut off parameter $\epsilon$, which might be due to the
large supersymmetry.
Moreover, the Casimir energy does not depend on the indices
$I,J$, and this means that the Casimir energy is the same for all
choices of holonomy $(n_0, \cdots , n_{k-1})$. This is consistent
with the argument in \cite{LM} that the vacua with different holonomy
are degenerated at zero temperature.
Notice that the Casimir energy
\begin{align}
 V_0 = N^2 \frac{3}{16 k}
 \label{casimir0}
\end{align}
is precisely the same as the mass of $AdS_5/{\mathbb Z}_k$ 
as we will discuss below.

For the case with the anti-periodic boundary condition, 
the above cancellation 
among $I,J$ dependent terms does not occur in general. 
With $L = J - I$ mod $k$ $( 0 \leq L < k)$
we find 
\begin{align}
\begin{aligned}
 \sum_E  ( 6 n^{I,J}_{S}  + n^{I,J}_{V^+}
  & + n^{I,J}_{V^-} - 4 n^{I,J+ \frac{k}{2}}_{F^+} 
 - 4 n^{I,J+ \frac{k}{2}}_{F^-} )
  E e^{- \epsilon E} \\ & =
 \frac{3}{8 k} + \frac{k}{6} -\frac{k^3}{24}
    + L^2 \left[ k - \frac{2}{3} \left( 2 L + \frac{1}{L} \right) \right]
    + {\cal O}(\epsilon) 
 \end{aligned}
 \label{nonzero}
\end{align}
for $0 \leq L \leq \frac{k}{2}$ and $L$ is replaced
by $k - L$  for $ \frac{k}{2} < L < k$.
Notice that the divergent terms proportional to 
$1/\epsilon^4$ and $1/\epsilon^2$ cancel out even in 
this case. 
The Casimir energy depends on the choice of holonomy
due to the $L$-dependence of the zero point energy \eqref{nonzero}.
The dominant contribution to the total partition function
comes from the vacuum with smallest zero point energy, 
which is realized  with the trivial holonomy. 
This is because the
value inside the bracket in \eqref{nonzero} is 
always positive for all $0 \leq L \leq \frac{k}{2}$
if we set $k \geq 4$.%
\footnote{For $k=2$ the cases with the periodic and 
anti-periodic boundary conditions lead to the identical result.}
The Casimir energy for the trivial holonomy case is given by
\begin{align}
 V_0 = N^2 \left( \frac{3}{16 k} + \frac{k}{12}-\frac{k^3}{48}  \right)~,
\label{casimir1}
\end{align}
which will be compared with the mass of the dual geometry \eqref{Mdf}.
Another interesting case may be with the ${\mathbb Z}_k$
symmetric holonomy. In this case, we sum up every $L$ with
the same weight, and hence we except that the cancellation
between the sectors with $L$ and $L + k/2$ occurs.
This can be confirmed by a direct computation, and the
Casimir energy is obtained as \eqref{casimir0}.%
\footnote{The Casimir energy in this case was already computed in 
\cite{CM1,CM2} by following the general method of \cite{BD}.}

Before finishing the arguments on Casimir energy,
we would like to make a comment on the validity of 
regularization adopted here, even though the divergent
terms cancel out in the final expressions.
Let us write the radius $R$ of $S^3$ explicitly 
such as  $E = (j+1)/R$ for the scalar case.
Then the divergent parts are proportional to $\mu^4 R^3 $ and
$\mu^2 R $ in each single-particle partition function. 
For quantum field theory on a curved background,
the divergent terms of energy momentum tensor should
be absorbed by the renormalization of coefficients
in the Einstein-Hilbert action. In our case
the divergent terms may be absorbed by
the counter terms $a \mu^4 \int \sqrt g$ and 
$b \mu^2 \int \sqrt g {\cal R}$. 
See \cite{BD} for more detailed discussions.

\subsection{Just above the critical temperatures}
\label{critical}

The eigenvalues distribute uniformly due to the repulsive potential
at low temperature, however the eigenvalues get together due to 
the attractive potential above the critical temperature. 
In particular, the densities may be gaped and vanish except for 
$- \theta_{I c} \leq \theta_I \leq \theta_{I c}$.
The condition that an eigenvalue $\theta_I$ does not feel any 
force is obtained as
\begin{align}
 n_I^2 \int d \theta ' _I \rho^I (\theta ' _I ) 
  \cot \left( \frac{ \theta_I - \theta ' _I}{2} \right)
  = 2 \sum_{J = 0}^{k - 1} n_I n_J \sum_{n=1}^{\infty} z^{I,J}_n (x)
     \rho_n^J \sin ( n \theta_I ) 
    \label{cond}
\end{align}
from the action \eqref{eff2}.
The general solutions subject to the normalization condition
$\rho_0^I = 1$
can be obtained by following \cite{JZ,AMMPR} in principle.
However, the analysis is quite complicated generically, so 
we adopt an approximation by setting 
$z^{I,J}_n (x) = 0$ for $n > 1$.
This approximation may be justified for small $x \sim x_c$ as
in Table \ref{table} by the fact that $z_n^{I,J}(x)$ with 
$n > 1$ is much smaller than $z_1^{I,J}(x)$.
In the following we will explicitly solve these equations
for the trivial holonomy case with $V=1$
and the ${\mathbb Z}_k$ symmetric case with $n_I = N/k$.

\begin{figure}[htbp]
 \begin{minipage}{0.5\hsize}
  \begin{center}
   \includegraphics[width=70mm]{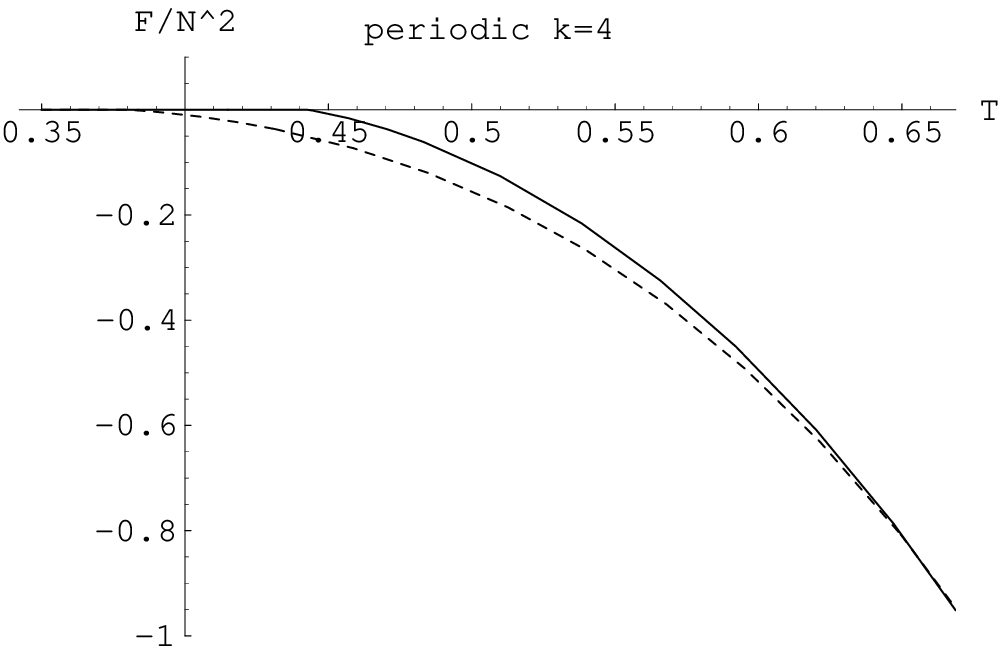}
  \end{center}
 \end{minipage}
 \begin{minipage}{0.5\hsize}
  \begin{center}
   \includegraphics[width=70mm]{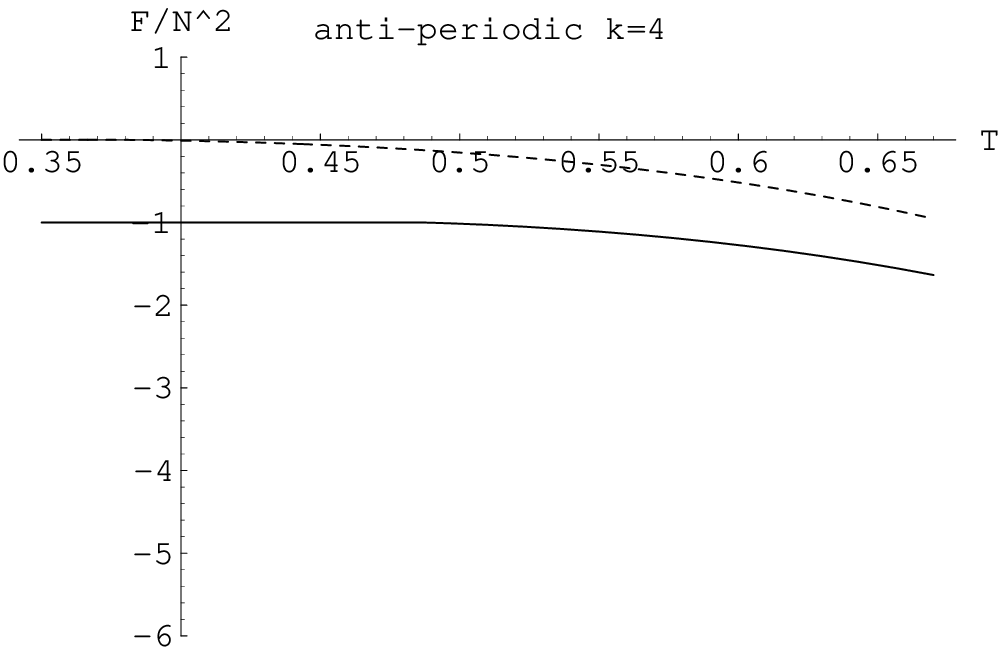}
  \end{center}
 \end{minipage} \\ 
 \begin{minipage}{0.5\hsize}
  \begin{center}
   \includegraphics[width=70mm]{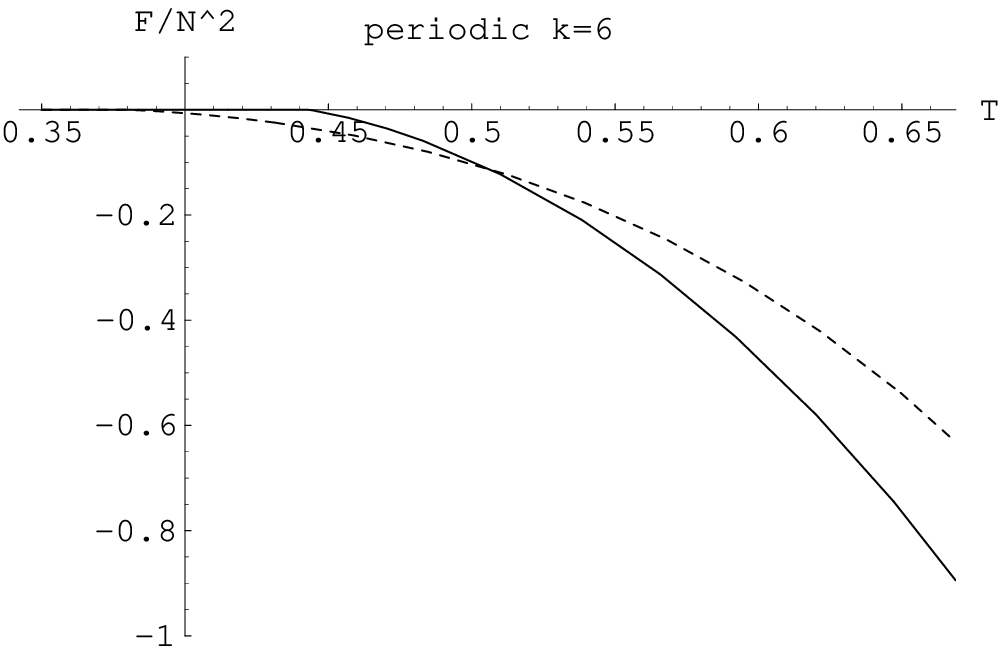}
  \end{center}
 \end{minipage}
 \begin{minipage}{0.5\hsize}
  \begin{center}
   \includegraphics[width=70mm]{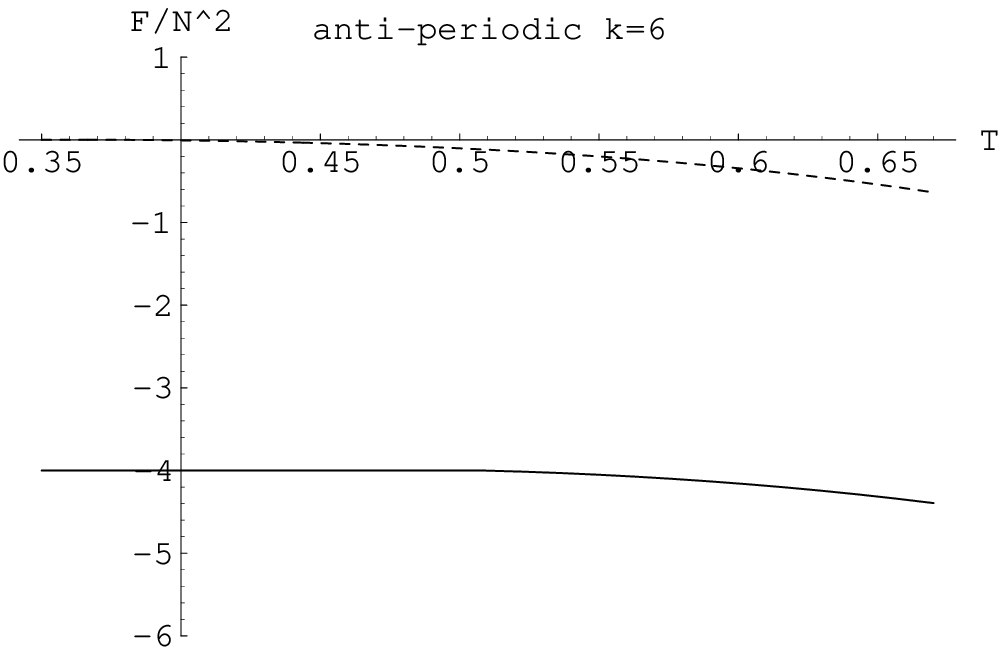}
  \end{center}
 \end{minipage}
  \caption{\it Free energies $F(T)/N^2$ as functions of $T$ in the
               cases with the periodic and the anti-periodic boundary
               conditions and with $k=4,6$. 
               The solid lines are for the trivial holonomy case
               and the dotted lines are for the ${\mathbb Z}_k$ symmetric case.}
  \label{free}
\end{figure}

Let us begin with the trivial holonomy case, where
the condition \eqref{cond} reduces to
\begin{align}
 \int d \theta_0 ' \rho^0 (\theta_0 ' ) 
  \cot \left( \frac{ \theta_0 - \theta_0 '}{2} \right)
  = 2 z^{0,0}_1 (x) 
     \rho^0_1 \sin \theta_0   ~.
\end{align}
This case is almost the same as the ${\mathbb R} \times S^3$ case analyzed in 
\cite{Sundborg,AMMPR}.
The solution is given by the form of
the Gross-Witten ansatz \cite{GW} as
\begin{align}
 \rho^0 (\theta_0 ) = \frac{1}{\pi \sin^2 \frac{\theta_c}{2}}
  \sqrt{\sin^2 \frac{\theta_c}{2} - \sin^2 \frac{\theta_0}{2}}
   \cos \frac{\theta_0}{2}
\end{align}
for $- \theta_c \leq \theta_0 \leq \theta_c $ and zero for otherwise.
The parameter $\theta_c$ satisfies
\begin{align}
 \sin ^2 \frac{\theta_c}{2} = 1 - \sqrt{1 - \frac{1}{z^{0,0}_1 (x)}} ~.
\end{align}
With this solution we can compute the classical action and
the free energy $F = - T \ln Z = T \langle S \rangle$ as
\begin{align}
 \frac{F}{N^2} \simeq
 V_0 - T \left( \frac{1}{2 \sin^2 \frac{\theta_c}{2}} + \frac12 
 \ln \sin^2 \frac{\theta_c}{2} - \frac12 \right) ~.
\end{align}
Near the critical temperature, it is given as
\begin{align}
  \frac{F}{N^2} \simeq V_0 - 
 \frac{T_H}{4} ( T - T_c ) \frac{ d}{d T} z^{0,0}_1 (e^{ - 1/T})|_{T = T_c}  
 + {\cal O} \left( (T - T_c)^2 \right) ~.
 \label{neartriv}
\end{align}
For the purpose of comparison with the ${\mathbb Z}_k$ symmetric case, 
we draw plots of the free energies for $k=4,6$ in Figure \ref{free}.
In the Figure we have shifted the zero point energy by
$3/(16 k)$.

For the ${\mathbb Z}_k$ symmetric case, the condition
\eqref{cond} becomes
\begin{align}
 \int d \theta ' _I \rho^I (\theta ' _I ) 
  \cot \left( \frac{ \theta_I - \theta ' _I}{2} \right)
  = 2 \sum_{J = 0}^{k - 1}  z^{I,J}_1 (x)
    \rho_n^J \sin  \theta_I  ~,
\end{align}
and the generic solutions are quite complicated.
However we only need the solution
responsible to the phase transition at the critical
temperature $x_c$ satisfying $\sum_{J} z_1^{0,J}(x_c) = 1$.
With the help of the ${\mathbb Z}_k$ symmetry,
we assign that the densities of eigenvalue take the same form as
$\rho^I (\theta^I) = \rho (\theta^I)$ for all $I$.
Then the solution can be easily found as
\begin{align}
 \rho (\theta_I ) = \frac{1}{\pi \sin^2 \frac{\theta_c}{2}}
  \sqrt{\sin^2 \frac{\theta_c}{2} - \sin^2 \frac{\theta_I}{2}}
   \cos \frac{\theta}{2}
\end{align}
for $- \theta_c \leq \theta \leq \theta_c $ and zero for otherwise.
In this case $\theta_c$ satisfies
\begin{align}
 \sin ^2 \frac{\theta_c}{2} = 
1 - \sqrt{1 - \frac{1}{\sum_{J=0}^{k-1} z^{0,J}_1 (x)}} ~.
\end{align}
This is valid for $x$ satisfying
$\sum_{J} z_1^{0,J} (x) \geq 1$,
and the equality
holds for the critical temperature $x = x_c$.
The free energy is
\begin{align}
 \frac{F^2}{N^2} \simeq
 V_0 - \frac{T}{k} \left( \frac{1}{2 \sin^2 \frac{\theta_c}{2}} 
 + \frac12 \ln \sin^2 \frac{\theta_c}{2} - \frac12 \right) ~,
\end{align}
and near the critical temperature
\begin{align}
  \frac{F^2}{N^2} \simeq V_0 - 
 \frac{T_H}{4} ( T - T_c ) \frac{d}{d T} \frac{1}{k}
\sum_{J=0}^{k-1} z^{0,J}_1 (e^{ - 1/T})  |_{ T = T_c}
 + {\cal O} \left( (T - T_c)^2 \right) ~.
 \label{nearsym}
\end{align}
See Figures \ref{free} for $k = 4,6$.

Let us compare the free energies for the above two cases.
For the case with the periodic boundary condition, 
the free energy for the ${\mathbb Z}_k$ symmetric case
is lower near the critical temperature,
since the critical temperature is smaller in this case.
Near the critical temperature, the free energy is
proportional to $T - T_c$ as in \eqref{neartriv} and 
\eqref{nearsym}, and the coefficients in \eqref{nearsym}
behaves like $1/k$.%
\footnote{This is checked for small $k$ by a numerical computation.} 
{}From this reason the free energy for the trivial holonomy case is
lower at slightly higher temperature for large $k$ 
as in Figure \ref{free}.
For the case with the anti-periodic boundary condition, 
the free energy for the trivial holonomy case is lower than for the 
${\mathbb Z}_k$ symmetric case due to the Casimir energy
$V_0$. Since the Casimir energy behaves like $k^3$,
the difference of free energy becomes bigger for larger $k$.

\subsection{High temperature behaviors}
\label{high}

At higher temperature the approximation in the previous subsection
is not valid any more, and the contribution from $n > 1$ terms
in \eqref{cond} should be taken into account.
Fortunately, the analysis becomes simpler when we take 
the high temperature limit $T \gg 1$.
This limit is the same as the limit of large radius $R$ of $S^3$,
where we can use the flat space approximation.
In this limit, the densities of eigenvalue may be given by
the delta function, thus we set $\rho^I_n = 1$ for all $n$. 
{}From \eqref{V^L} we find
\begin{align}
\begin{aligned}
 \sum_{n=1}^{\infty} V_n^L
  = - \zeta (4) T^3 
  \left[ \frac{16}{k} + \left( 1 - \frac{1}{2^3} \right) \frac{16}{k} \right]
  - \zeta (2) T 
  \left[ - \frac{2}{k} + \left( 1 - \frac{1}{2} \right) \left(- \frac{2}{k}\right ) \right]
  + {\cal O} \left(\frac{1}{T}\right) ~,
\end{aligned}
\end{align}
which may be read off from the expansion of 
$z^{I,J}_A (e^{- \epsilon})$ 
by $\epsilon$.
The coefficients of $T^3$ terms can be given by the degrees 
of freedom in the flat space limit, where the volume should be
divided by $k$.
Since the above expression does not depend on $L$, the free energy
behaves in the same way for all the vacua and for both the
spin structures in the high temperature expansion.
This is quite natural since the high energy excitations 
should not depend on the vacuum structure.
In summary, the free energy $F = - T \ln Z$ 
and the expectation value of energy 
$E = - \frac{\partial}{\partial \beta} \ln Z$
are given by
\begin{align}
 F &= - \frac{N^2}{k} \left( \frac{\pi ^4}{3} T^4 - \frac{\pi ^2} {2} T^2 \right)
  + {\cal O}(1) ~,
 &E &= \frac{N^2}{k} \left ( \pi^4 T^4 - \frac{\pi^2}{2} T^2 \right)
    + {\cal O}(1)
\label{energy}
\end{align}
for every choice of holonomy and
for both the spin structures.
This energy was already computed in \cite{HJ} in a different way.


\section{Dual gravity description}
\label{gravity}

In the limit of large 't Hooft coupling the dual
gravity description is more appropriate to
discuss the phase structure of the gauge theories.
In the dual picture the confinement/deconfinement phase
transition is described by the Hawking-Page transition
\cite{HP,Wittent}.
In this section we investigate the Hawking-Page transition
in the dual geometries and try to see whether the phase 
structure continues from the one at zero 't Hooft coupling.
In the next subsection we review the
Hawking-Page transition between the thermal
AdS space and the AdS-Schwarzschild black hole.
In subsection \ref{hporbifold} we move to
our orbifold cases.
With the periodic boundary condition, 
the geometries are given by various arrangements
of $k$ NS5-branes in a T-dual picture.
With the anti-periodic boundary condition,
there are localized tachyons at the fixed
point of the thermal AdS orbifold. 
The condensation of the localized tachyon 
is discussed in subsection \ref{localized}.

\subsection{Hawking-Page transition}
\label{hp}

The boundary gauge theory at finite temperature may be
defined on $S^1 \times S^3$ with a thermal cycle,
and we have to include all geometries with the same
boundary condition to compute a path integral in gravity
theory. The geometries are obtained by extending the boundary
$S^1 \times S^3$ into the bulk, which can be done in two ways.
One of the geometry has the
topology of $S^1 \times B^4$, where the 4 dimensional ball $B^4$ 
is obtained by filling the inside of $S^3$. This geometry 
is the thermal AdS space, where the thermal cycle has a periodicity.
The other geometry has the topology $B^2 \times S^3$,
which is the AdS-Schwarzschild black hole.
We will see that the thermal AdS space is dominant at low
temperature and the AdS-Schwarzschild
black hole is dominant at high temperature.

The metric of the thermal $AdS_5$ is given by
\begin{align}
 ds^2 &= g(r) dt^2 + \frac{dr^2}{g(r)} + 
   r^2 d s^2_{S^3} ~,
  &g(r) &= \frac{r^2}{l^2}  + 1 ~,
\label{AdS}
\end{align}
where $d s^2_{S^3}$ is the metric of $S^3$ and the 
Euclidean time is periodic  $t \sim t + \beta_l $.
The mass of the thermal $AdS_5$ was computed in \cite{BK} as
\begin{align}
 M = \frac{3 \pi l^2}{32 G_5}
 \label{AdSmass}
\end{align}
by utilizing the boundary stress tensor method.
This is precisely the same as the Casimir energy of
${\cal N}=4$ $U(N)$ super Yang-Mills theory on $R \times S^3$ 
given in \eqref{casimir0} with 
$k=1$. The 5 dimensional Newton constant $G_5$ is written as
$G_5 = \pi l^3 / (2 N)$ in terms of the dual gauge theory.
For simplicity we set the AdS radius as $l=1$.

The metric of the AdS-Schwarzschild black hole is
\begin{align}
 ds^2 &= h(r) dt^2 + \frac{dr^2}{h(r)} + 
   r^2 d s^2_{S^3} ~,
  &h(r) &= r^2 + 1 - \frac{r_0^2}{r^2} ~,
  \label{AdSBH}
\end{align}
where the period of the Euclidean time is
given by $\beta_h = 2 \pi r_+ /(2 r_+^2 + 1)$.
We denote $r_+$ as the horizon satisfying 
$h (r_+) =0$ or equivalently $r^2_0 = r^2_+ + r^4_+ $.%
\footnote{There are two solutions $r_{\pm}$ to this equation,
which implies that  there are two types of black holes.
The bigger and smaller ones $r_+,r_-$ 
are the radii of horizons of big and small black holes, 
respectively.
We only consider the big black hole since
the small black hole has a negative specific heat and hence
it is unstable.
However the unstable saddle point may be 
important to understand the phase structure as pointed 
out in \cite{AMMPR,ALW,ABMW}.}
The Mass of the AdS-Schwarzschild black hole is \cite{BK}
\begin{align}
 M = \frac{3 \pi r_0^2}{8 G_5} + \frac{3 \pi }{32 G_5} ~,
 \label{BHmass}
\end{align}
where the second term is the constant $AdS_5$ contribution.
This mass can be expanded by $T = 1/\beta_h$ at high temperature as 
\cite{HJ}
\begin{align}
 M = \frac{3 \pi}{8 G_5} ( \pi^4 T^4 - \pi^2 T^2 ) + {\cal O}(1) ~,
\end{align}
which is about 3/4 times the energy given in \eqref{energy} with $k=1$.%
\footnote{It was argued in \cite{HJ} that the origin of the $3/4$
difference is the same as the one of \cite{GKP}, where the entropy
of black 3-branes is compared with the state counting on
D3-brane.}

Now we can compute the partition function in the gravity 
description. In the classical approximation, the partition 
function is obtained from the classical actions for the 
geometries, which are proportional to the volume as
$S = V/(2 \pi G_5)$. Since the volume diverges for both cases,
we introduce a cut off $r_m$ and examine the difference.
The Euclidean time periods are set equal at $r_m$ as
$\beta_l \sqrt{g(r_m)} = \beta_h \sqrt{h(r_m)}$.
Then the difference of the classical action is computed as
\cite{HP,Wittent}
\begin{align}
\lim_{r_m \to \infty} \frac{V_{BH} (r_m) - V_{TAdS} (r_m)}{2 \pi G_5}
= \frac{\pi^2 (r_+^3 - r_+^5)}{4 G_5 (2 r^2_+ + 1)} ~.
 \label{ptt}
\end{align}
For small $r_+$ (low temperature) the above quantity is positive,
which means that the thermal AdS space dominates the partition
function. The phase transition occurs at $r_+ =1$ or 
$T = 3/(2 \pi)$, and above the critical temperature
the AdS-Schwarzschild black hole dominates.

In the dual gauge theory, the Polyakov loop 
$\langle \Tr U \rangle = \langle \Tr P \exp{i \oint A_t} \rangle$
is an important order parameter. 
In the confinement phase, the ${\mathbb Z}_N$ symmetry of the 
theory is unbroken, and hence the Polyakov loop vanishes,
which is realized by uniformly distributed 
eigenvalues $\theta_i$. In the deconfinement phase, since the 
${\mathbb Z}_N$ symmetry is broken, the Polyakov
loop may have a non-trivial value and the eigenvalues are
distributed non-trivially.%
\footnote{Because the eigenvalues collapse in the ${\mathbb Z}_N$
symmetric way, the Polyakov loop may vanish after taking 
the average.
We may use $\langle | \Tr U | \rangle$ or 
$\langle \Tr U^2 \rangle$ to avoid this subtlety.}
The Polyakov (Wilson) loop along the path ${\cal C}$ may be computed
in the gravity side as $\exp ( - A )$, where $A$ represents 
the minimum area of the worldsheet with boundary ${\cal C}$
subject to a regularization \cite{Rey,Maldacenaw}.
For the thermal AdS space, the thermal cycle is not contractible, so the
area is infinite and this leads to the vanishing Polyakov loop.
For the AdS-Schwarzschild black hole, the thermal cycle shrinks
at the horizon, so the area could be finite and hence
the Polyakov loop can take a non-zero value.
See \cite{Wittent} for more detail.

\subsection{Phase transitions of the orbifolds}
\label{hporbifold}

Due to the ${\mathbb Z}_k$ symmetry
the vacuum with the holonomy $n_I = N/k$ for all $I$ 
is supposed to be dual to the standard orbifold.
At low temperature,
the dual geometry is given by the orbifold of the thermal AdS space,
whose metric is \eqref{AdS} with $ds^2_{S^3}$ replaced by
\begin{align}
 ds^2_{\rm orb} = \frac14 [
  ( d \chi + \cos \theta d \phi )^2 + d \theta ^2 
   + \sin ^2 \theta d \phi ^2 ]~,
   \label{metricorb}
\end{align}
where the variables run $0 \leq \theta \leq \pi$, 
$0 \leq \phi \leq 2 \pi$ and $0 \leq \chi \leq 4 \pi /k$
due to the orbifold identification.
This metric of $\sz$ may be useful since we can easily
see that the orbifold action acts on the $\chi$-cycle as
$\chi \to \chi + 4 \pi/k$ and that there is no fixed point
on this space.
The mass of the thermal AdS orbifold is 
\begin{align}
 M = \frac{3 \pi}{32 k G_5} ~,
 \label{AdSmassorb}
\end{align}
which is $1/k$ times the mass of the thermal AdS space \eqref{AdSmass}.
This should be compared with the Casimir energy with the
${\mathbb Z}_k$ symmetric holonomy. The Casimir energy 
is given by \eqref{casimir0} for both the spin structures
and it is precisely the same as \eqref{AdSmassorb}.
The geometry of $AdS_5$ is believed to be stable under
$\alpha '$ correction and the stability seems to 
continue to the standard orbifold.

The thermal AdS orbifold has a fixed point at $r=0$ and
there are closed strings in twisted sectors localized 
at the fixed point.
With the periodic boundary condition for the fermions along
the $\chi$-cycle, there are massless states localized at the
fixed point, and the different vacua are dual to different
excitations of these massless states.
The excitation of localized massless states does not change 
the global geometry, thus all the geometries contribute equally 
to the partition function. 
In particular, the mass of the every geometries should be 
the same as \eqref{AdSmassorb}, which is also the same as the 
Casimir energy for the every vacua.
Therefore, we can say that the phase structure does not 
change even at the large 't Hooft coupling.%
\footnote{We are not sure whether the phase structure is the same
in a middle value of the 't Hooft coupling, but we guess 
that this is indeed the case.}
For large $k$, the T-dual picture along the $\chi$-cycle is
relevant, where the $k$ NS5-branes are arranged in the dual
$\tilde \chi$-cycle \cite{OV}. In the T-dual picture,
the different excitations of massless
states correspond to different configurations of $k$ NS5-branes.%
\footnote{It was shown in \cite{LM} that the dual geometries
are also labelled by the integers $(n_0, \cdots , n_{k-1})$
at zero temperature. They discussed how to construct these
geometries, where the NS5-branes are replaced by flux and 
the back-reaction of the flux is taken into account.}
In particular the vacuum with the trivial holonomy corresponds to
the configuration of $k$ coincident NS5-branes.
With the anti-periodic boundary condition, there are localized 
tachyons at the fixed point as in \cite{APS}.
The different vacua are dual to different condensations
of these localized tachyons, which deform
the geometry significantly from the thermal AdS orbifold.
Since the configurations with tachyon are unstable,
the relevant geometry should be the one without tachyon, 
which will be discussed in the next subsection.

At high temperature, the dual geometry is the orbifold
of the AdS-Schwarzschild black hole \eqref{AdSBH} with
$ds^2_{S^3}$ replaced by \eqref{metricorb}.
The mass of the black hole may be given by the expansion of the
temperature as
\begin{align}
 M = \frac{3 \pi}{8 k G_5} ( \pi^4 T^4 - \pi^2 T^2 ) + {\cal O}(1) ~,
\end{align}
which is $3/4$ times the energy given in \eqref{energy}
as discussed in \cite{HJ}.
Since the geometry does not have any fixed point,
there are no light localized modes generically.
Due to this fact, the orbifold of the black hole seems to be 
the most relevant one, even though the every vacua are degenerated
at the zero 't Hooft coupling. 
At relatively small temperature or for large $k$, there may 
be nearly massless modes or tachyonic modes near the horizon.
With the periodic boundary condition, the T-dual picture is
more relevant for large $k$, where the shift of NS5-branes
is given by the nearly massless modes.
With the anti-periodic boundary condition, the orbifold
of the AdS-Schwarzschild black hole may decay into a
resolved AdS orbifold through the tachyon condensation.
This type of geometry decay has been discussed in \cite{Horowitz,HS}.
According to them, a kind of black hole decays into a bubble 
of noting by a winding tachyon condensation.

\subsection{Localized tachyon condensation}
\label{localized}

If the anti-periodic boundary condition is assigned
for the fermions along the $\chi$-cycle, then
there are tachyonic modes at the fix point of the
thermal AdS orbifold at low temperature. 
The geometry deformed by a tachyon condensation
was proposed by \cite{CM1,CM2} as%
\footnote{See also \cite{nuttier}.}
\begin{align}
 ds^2 = g(r) dt^2 + \frac{dr^2}{g(r)f(r)}
      + \frac{r^2}{4}
       [ f(r) ( d \chi + \cos \theta d \phi )^2 + d \theta ^2
            + \sin ^2 \theta d \phi ^2 ] ~,
            \label{EH}
\end{align}
with
\begin{align}
  g(r) &= r^2  + 1 ~,
&f(r) &= 1 - \frac{a^4}{r^4} ~, 
&a^2 &= \left(\frac{ k^2}{4} - 1\right) ~.
\label{defa}
\end{align}
They call this geometry as Eguchi-Hanson soliton.
We set $k$ even in order to assign the anti-periodic 
boundary condition for the fermions and also
$k > 2$ for $a^2 > 0$.
Because of the boundary condition for the fermions,
the non-trivial $\chi$-cycle can be pinched off.
The boundary metric at $r \to \infty$ is given by
$\eqref{metricorb}$ with $\chi \sim \chi + 4\pi/k$
as supposed to be.
The solution is regular everywhere including $r=a$,
and there are no tachyonic modes.
There is no geometry for $r < a$, and this region may 
be replaced by the tachyon state or the nothing state in the sense 
of \cite{HS}.

The conserved mass of the deformed geometry was computed in 
\cite{CM1,CM2} as
\begin{align}
 M =\frac{\pi (3 - 4 a^2)}{32 k G_5} = - \frac{\pi (k^4 - 8 k^2 + 4)}{128 k G_5}
  ~,
 \label{Mdf}
\end{align}
where $G_5 = \pi/(2 N^2)$ in the gauge theory terminology.
Since the non-trivial $\chi$-cycle is pinched off at $r=a$, 
the Wilson loop along the $\chi$-cycle can take non-trivial 
value according to the previous discussion on the Polyakov loop. 
This means that the ${\mathbb Z}_k$ symmetry is broken in this 
background, and this is consistent with the fact that the dual 
vacuum has the trivial holonomy.
It is amusing to notice that the mass of the Eguchi-Hanson soliton 
\eqref{Mdf} is about $3/4$ times the Casimir energy of the
gauge theory on the trivial holonomy vacuum \eqref{casimir1}
for large $k$.
This reminds us of the famous $3/4$ difference in the context
of \cite{GKP} mentioned above, but their origins are not 
directly related to each other.
It is worth to study this issue furthermore.

At enough high temperature, only the relevant geometry
seems to be the orbifold of the AdS-Schwarzschild black hole.
Therefore we may observe the phase transition between
the Eguchi-Hanson soliton and the
orbifold of the AdS-Schwarzschild black hole.
As in the previous case the partition
function may be obtained from the classical actions
for these geometries in the classical approximation.
The difference of these actions is given by
\begin{align}
\lim_{r_m \to \infty} \frac{V_{BH} (r_m) - V_{EH} (r_m)}{2 \pi G_5}
= \frac{\pi^2 r^+ (r_+^2 - r_+^4 - 2 a^2)}{4 k G_5 (2 r^2_+ + 1)} ~,
\label{pttorb}
\end{align}
where the region of $r < a$ is removed in the Eguchi-Hanson soliton.
The critical temperature is%
\footnote{Notice that the critical temperature of the 
Hawking-Page transition $T_c = 3/(2 \pi)$ is reproduced 
for $a=0$.}
\begin{align}
 T_c = \frac{2 + \sqrt{1 + 8 a^2}}{\sqrt{2 \pi^2 ( 1 + \sqrt{1 + 8 a^2})}} ~.
\end{align}
We can see from \eqref{pttorb} that the Eguchi-Hanson soliton is
dominant at lower temperature and the orbifold of AdS-Schwarzschild
black hole is dominant at higher temperature.

\section{Conclusion and discussions}
\label{conclusion}

We have studied the thermodynamics of ${\cal N}=4$ $U(N)$ super 
Yang-Mills theories on ${\mathbb R} \times S^3/{\mathbb Z}_k$ with large $N$.
The base manifold $S^3/{\mathbb Z}_k$
has a non-trivial cycle along the $\chi$-direction of
\eqref{metricorb}, and a non-trivial holonomy can be assigned
along the non-trivial cycle. 
The theory has multi-vacua
associated with the choice of holonomy labelled by
$k$ integer numbers $(n_0 , \cdots , n_{k-1})$.
We can assign the periodic and anti-periodic boundary conditions
for the fermions along the non-trivial cycle.
On a compact manifold,
the Gauss constraint only allows gauge invariant operators,
and due to this fact a phase transition occurs even in the
zero 't Hooft coupling limit for large $N$.
We have computed the partition function for the large $N$ gauge
theories with different holonomies by following the analysis in
\cite{Sundborg,AMMPR}, and examined the phase structure
with a special care on the difference between the vacua.

At low temperature, the most relevant contribution to the
free energy comes form the Casimir energy.
For the case with the periodic boundary condition,
the Casimir energy does not depend on the choice of holonomy, 
and hence the vacua are degenerated.
For the case with the anti-periodic boundary condition, 
the Casimir energy depends on the choice of holonomy, and 
the dominant contribution comes from the vacuum with the
trivial holonomy $V=1$.
Near the critical temperature, we can obtain an approximate 
analytic expression of free energy by utilizing the 
Gross-Witten ansatz \cite{GW}. See Figure \ref{free}.
For enough large $k$ the case with the trivial holonomy
seems to dominate for both the spin structures.
At high temperature, the free energy is universal as in
\eqref{energy}, thus the vacua are degenerated for 
both the spin structures.%
\footnote{The masses of lightest twisted string modes are
proportional to the horizon radius $r_+$, and these modes 
become massless (or tachyonic) in the zero 't Hooft 
coupling limit since the radius behaves like 
$r_+ \sim \alpha ' \sqrt{\lambda} T$.
{}From this reason we can assign non-trivial expectation 
values to the lightest modes and the vacua can be 
degenerated in the gauge theory description.}

In the limit of large 't Hooft coupling, the dual gravity 
description is more appropriate.
In the low temperature phase, the dual geometry is the 
orbifold of the thermal AdS space or its deformation. 
The orbifold of the thermal AdS space has the fixed point at $r=0$,
and there are closed string states localized at the fixed point.
With the periodic boundary condition, the localized states
are massless, and the excitation of these massless states
leads to degenerated different geometries. 
In a T-dual picture, the different excitations correspond to 
different arrangements of $k$ NS5-branes along 
the dual $\tilde \chi$-cycle. 
With the anti-periodic boundary condition, the localized states
are tachyonic and a condensation of the tachyonic modes 
leads to the decay of geometry into the regularized geometry 
\eqref{EH} called as the Eguchi-Hanson soliton \cite{CM1,CM2}.
In the viewpoint of the dual gauge theory, the localized tachyon 
condensation is realized as the transition between different
vacua. In particular, we have found that the mass of the 
Eguchi-Hanson soliton is about $3/4$ times the Casimir energy 
for the dual vacuum with the trivial holonomy.
I would like to study the relation between the vacuum transition
of the gauge theory and the RG-flow or time-dependent 
process among the geometries as in \cite{APS}.
At enough high temperature, the dual geometry is the
orbifold of the AdS-Schwarzschild black hole. Since there is
no fixed point in this geometry, the orbifold 
seems to be the most relevant geometry. 
This implies that the phase structure would vary as the 't Hooft 
coupling is changed at high temperature.

There are several theories similar to our orbifold gauge theories
in the sense that the gauge theory has many vacua and its dual
gravity description.
One of them is (1+1) dimensional large $N$ gauge theories on a 
torus \cite{GL1,GL2}, where non-trivial holonomy matrices can be
assigned along the two cycles.
For the case with the periodic boundary condition of the 
fermions along the spatial cycle, the eigenvalues of spatial 
holonomy matrix correspond to the positions of $N$ D0-branes along 
the T-dual spatial cycle \cite{GL1,GL2}.
In the high temperature phase the eigenvalues are distributed 
uniformly, but in the low temperature phase the eigenvalues 
get together.
This is related to the Gregory-Laflamme transition from a black 
string wrapped on a spatial cycle into a localized black hole \cite{GL}.
For the case with the anti-periodic boundary condition, 
the spatial cycle can shrink, and indeed the thermal 
$AdS_3$ is dominant in the low temperature phase. 
In this phase, the ${\mathbb Z}_N$ symmetry of the holonomy 
matrix along the spatial cycle has to be broken.
In the high temperature phase, the BTZ black hole dominates, 
and the ${\mathbb Z}_N$ symmetry is preserved.
The relation to our cases may be examined for large $k$ limit,
where the back-reaction of NS5-branes should be taken into account.
I would like to study this relation, for instance, by taking
a large $N,k$ limit with keeping a ratio $N/k$ finite.%
\footnote{
A similar limit was taken in \cite{Schnitze} for the case
including $N_f$ fundamental matters with a finite {$N_f/N$},
where they observed a third order phase transition just like
the Gross-Witten transition \cite{GW}.}

Other interesting theories are the plane wave matrix model
\cite{BMN} and (2+1) dimensional super Yang-Mills theory on 
${\mathbb R} \times S^2$,
which are obtained by truncating the
${\cal N}=4$ Yang-Mills theory on ${\mathbb R} \times S^3$
like our orbifold theory.
These models share the same symmetry $SU(2|4)$ at zero
temperature, and the gravity dual of these models was
studied in \cite{LM}.
The thermodynamics of plane wave matrix model has been 
studied in \cite{FSS,Semenoff,SRV,HRSY,KNY}, and,
in particular, the different vacua was compared in \cite{KNY}.
The thermodynamics of large $N$ gauge theory on 
${\mathbb R} \times S^2$ should be also interesting.
The relation among these models at zero temperature
has been discussed in \cite{LM,Ling,Lin,ISTT},
and it is worth while investigating the relation
among their thermodynamics.

In this paper, we have taken the zero 't Hooft coupling
limit $\lambda = 0$, namely, the free theory limit, 
thus a next task is to study the effects of non-zero 
't Hooft coupling.
In the free theory limit, the phase transition is of the first
order, but the inclusion of small $\lambda$ may change the
order of phase transition as discussed in \cite{AMMPR,AMMPR2}.
Moreover, we may be able to examine how the phase structure in the 
zero 't Hooft coupling limit continues to the one in the
large 't Hooft coupling limit.%
\footnote{It might be useful to make use of the Penrose limit \cite{BMN}
to examine the intermediate regime. 
See \cite{HO,HO2} for the comparison of the Hagedorn temperature.}
For example, the Casimir energy on the trivial vacuum with
the anti-periodic boundary condition \eqref{casimir1} is 
about $4/3$ times the mass of the Eguchi-Hanson soliton \eqref{Mdf},
thus one may wonder what would happen if we include
$\lambda$ correction.
In the high temperature phase, we have observed that the relevant
geometry is the orbifold of the AdS-Schwarzschild black
hole, which is dual to the ${\mathbb Z}_k$ symmetric vacuum.
Since the vacua are degenerated in the zero 't Hooft coupling limit,
the vacuum structure should depend on the 't Hooft coupling.
We would like to study this issue as a future work.

\subsection*{Acknowledgement}

We would like to thank Y.~Nakayama, I.~Papadimitriou,
S.-J.~Rey, V.~Schomerus and K.~Yoshida for useful discussions
and Albert Einstein Institute for hospitality.
This work is supported by JSPS Postdoctoral Fellowships for 
Research Abroad H18-143.


\appendix

\section{The partition function of single scalar particle}
\label{example}

As an example
we compute here the partition function of single scalar particle.
Let us define the following function on $S^3$ as
\begin{align}
 f(x,y) = \Tr x^E y^{2 j_2} ~.
\end{align}
Since a scalar is expanded by the scalar spherical harmonics 
$S_{j,m,\bar m}(\Omega)$,
the above function can be easily evaluated as
\begin{align}
 f(x,y) = x + 2 x^2 ( y + 1/y) + 3 x^3 (y^2 + 1 + 1/y^2 )
 + \cdots .
\end{align}
The orbifold case is given by restricting the modes into
the ones invariant under the orbifold action \eqref{orbifolda}.

In the case of the trivial holonomy $V=1$, we have to project
the modes into the ones with $2 \bar m = k {\mathbb Z}$.
The projection leads to
\begin{align}
 f(x , y ) = x + 3 x^3 + \cdots + 
 \sum_{n = 1}^{\infty} ( y^{ n k } + 1/y^{ n k} ) 
 [ ( n k + 1) x^{n k + 1} + ( n k + 3 ) x^{n k + 3} + \cdots ] ~,
\end{align}
where only terms proportional to $y^{k {\mathbb Z} }$ are kept.
Using a formula
\begin{align}
 L x^L + ( L + 2) x^{L+2} + \cdots 
  = \frac{L x^{L} - ( L- 2 ) x^{L + 2}}{(1 - x^2 ) ^2} ~,
  \label{formula1}
\end{align}
the function can be written as 
\begin{align}
 f(x , y ) = \frac{x + x^3}{(1 - x)^2} + 
 \sum_{n = 1}^{\infty} ( y^{ n k } + 1/y^{ n k} ) 
  \frac{( n k + 1) x^{n k + 1} - ( n k - 1 ) x^{n k + 3}}{(1 - x^2)^2 } ~. 
\end{align}
With $\sum_{n=1}^{\infty} x^n = x/(1 - x)$ and
$\sum_{n=1}^{\infty} n x^n = x/(1 - x)^2$, we finally find for $y=1$
\begin{align}
 f(x , 1 ) = \frac{(x + x^3)(1 - x^{2 k}) + 2 k x^k (x - x^3)}{(1 - x)^2 ( 1 - x^k )^2} ~. 
\label{zls0}
\end{align}
This gives the expression in \eqref{zls} with $I=J$.

The case with non-trivial holonomy can be analyzed in a similar way.
For a bi-fundamental scalar $(n_I , \bar n_J)$,
the orbifold action is given by \eqref{orbifolda}.
In this case we keep the terms proportional to
$y^{k {\mathbb Z} + L}$, and we find
\begin{align}
\begin{aligned}
 f(x , y ) &=
  \sum_{n = 0}^{\infty}  
  y^{ n k + L } [ ( n k + L + 1 ) x^{n k + L + 1}
  +  ( n k + L + 3 ) x^{n k + L + 3} + \cdots ] 
 \\ &+
 \sum_{n = 1}^{\infty}  
  y^{ n k - L } [ ( n k - L + 1 ) x^{n k - L + 1}
  +  ( n k - L + 3 ) x^{n k - L + 3} + \cdots ] ~.
\end{aligned}
\end{align}
Making use of the formula \eqref{formula1},
we can reduce the above sum into the more simplified
form in \eqref{zls}.


\section{An index for the supersymmetric orbifold theory}
\label{index}

In this appendix the index proposed in \cite{KMMR} is 
computed for ${\cal N}=4$ $U(N)$
super Yang-Mills theory on $\rs$ with the periodic boundary condition
for the fermions along the $\chi$-cycle.%
\footnote{See also \cite{Romelsberger,Nakayama1,Nakayama2}.}
The index in our case is defined as
\begin{align}
 {\cal I} &= \Tr (-1)^F e^{- \beta \Xi} t^{2 ( E + j_1) } v^{R_2} w^{R_3} ~,
   &\Xi &= E - 2 j_1 - \frac{3}{2} R_1 - R_2 - \frac{1}{2} R_3 ~.
\end{align}
Since only states satisfying $\Xi = 0$ contribute the index \cite{KMMR},
we can set $\beta \to \infty$.
As before $E$ is the energy and $(j_1 , j_2)$ are spins with respect
to $SU(2)_1 \times SU(2)_2$ isometry on $S^3$.
The symmetry of the theory is $SU(2|4)$ and $R_1,R_2,R_3$ are
$R$-charges. 
With generic holonomy $(n_0,\cdots,n_{k-1})$, the $U(N)$ gauge 
symmetry is reduced to $\prod_I U(n_I)$, and the states are in
adjoint or bi-fundamental representation of the gauge group.
As in \cite{KMMR,Nakayama1} or in \eqref{pf2} the index can be 
written by an integral of unitary matrices as
\begin{align}
 {\cal I} = \int [ \prod_I d U_I ] \exp { \left[ \sum_{I,J} \sum_{n=1}^{\infty} 
 \frac{1}{n} f^{I,J} (t^n , v^n , w^n ) \Tr (U_I^n) \Tr (U^{\dagger n}_J) \right] } ~,
 \label{indexp}
\end{align}
where the coefficients $f^{I,J}$ are the indices for the single 
particles with representation $(n_I , \bar n_J)$.
Notice that the index does not receive any corrections of
 't Hooft coupling,
since we are counting the modes protected by supersymmetry 
\cite{KMMR}.

Let us compute the indices for single particles.
We start from the scalar contribution.
The theory includes three complex scalars $X$,$Y $,$ Z$ with 
$(R_1,R_2,R_2) = (0,1,0)$, $(1,-1,1)$, $(1,0,-1)$. 
The scalars are expanded by
$S_{j,m , \bar m} (\Omega)$, and the condition $\Xi = 0$ is satisfied
by the modes $S_{j,j/2,\bar m}$. The orbifold projection
means $2 \bar m \in L + k {\mathbb Z}$ for the $L = J - I + k {\mathbb Z}$
$(0 \leq L < k)$ sector, and the contribution can be computed as
\begin{align}
 \frac{t^2 ( t^{3 L }  + t^{3 ( k - L ) } ) ( v + 1/w + w/v)}
      {(1 - t^6)(1 - t^{3 k} )} ~.
\end{align}
Since the gauge field does not have any $R$-charges, the condition
$\Xi = 0$ is satisfied by the modes $V^+_{j,(j+1)/2,\bar m}$
with $ 2 \bar m \in L + k {\mathbb Z}$. Thus the contribution from
the gauge field is
\begin{align}
 \frac{t^6 ( t^{3 L }  + t^{3 ( k - L ) } )}
      {(1 - t^6)(1 - t^{3 k} )} ~.
\end{align}
For the fermions,
the condition
$\Xi = 0$ can be satisfied by the modes $F^{+}_{j,j/2,\bar m}$ with
$ 2 \bar m \in L + k {\mathbb Z}$ and
$(R_1,R_2,R_2) = (1,-1,0)$, $(0,1,-1)$, $(0,0,-1)$. 
The contribution from these fermions is
\begin{align}
 - \frac{ t ^4 ( t^{3 L }  + t^{3 ( k - L ) } ) ( 1/v + w + v/w)}
      {(1 - t^6)(1 - t^{3 k} )} ~.
\end{align}
The other contribution comes from the modes 
$F^{-}_{j,(j-1)/2,\bar m}$ with
$ 2 \bar m \in L + k {\mathbb Z}$ and
$(R_1,R_2,R_2) = (1,0,0)$ as
\begin{align}
 \delta_{L,0} - \frac{( t^{3 L}  + t^{3 ( k - L ) } )}
      {(1 - t^6)(1 - t^{3 k} )} ~.
\end{align}
Interestingly the sum of all contributions can be factorized
as
\begin{align}
  f^{I,J}(t,v,w)
 = \delta_{L,0} - \frac{(t^{3 L} + t^{3 (k-L)})(1 - t^2 v )(1-t^2 /w)(1 - t^2 w/ v)}{(1-t^6)(1-t^{3k})} ~.
\end{align}
For the indices to converge, we have to assign $t^2 < 1$,
$t^2 v < 1$, $t^2 /w < 1$, $t^2 w / v < 1$, which implies
$\delta_{I,J} - f^{I,J} (t,v,w) > 0$.

In order to perform the integral \eqref{indexp}, we assume that
$n_I$ is very large or zero. Then we can replace the discrete
eigenvalues by continuous ones with the densities
$\rho^I (\theta_I )$ satisfying $\int d \theta_I \rho^I (\theta_I) = 1$.
In this term, the effective action is given by 
\begin{align}
 S (\rho^I_n) &= \sum_{I,J} n_I n_J 
 \sum_{n = 1} \rho^I_n \bar \rho^J_n V^{I,J}_n ~,&
 V_n^{I,J} &= \frac{1}{n} (\delta_{I,J} - f^{I,J} (t^n, v^n ,w^n) ) ~,
\end{align}
where we denote the Fourier transform of $\rho^I(\theta_I)$ as $\rho^I_n$.
The saddle point of the action is
$\rho_n^I = 0$ for $n \neq 0$ as $V_n^{I,J} > 0$, and the index
is given by the determinant
\begin{align}
 {\cal I} = \prod_n \frac{1}{\det ( \sum_{I,J} n_I n_J V^{I,J}_n )} ~.
\end{align}
The determinant is complicated in general, but it could 
be written in a simpler form for several cases.

One case is with the trivial holonomy $V=1$. In this case, we have just
$1 \times 1$ matrices, so the determinant is simply
\begin{align}
 {\cal I} = \prod_n  
 \frac{(1-t^{ 6 n} )(1-t^{3k n})}
  {( 1 + t^{3k n})(1 - t^{2 n} v^n )(1-t^{2 n}/w^n)
   (1 - t^{2 n} w^n / v^n )}~. 
   \label{indextriv}
\end{align}
The $N$ dependent factor is removed by changing the normalization.
Another interesting case may be with the ${\mathbb Z}_k$ symmetric
holonomy $n_I = N/k$ for all $I$.
In this case, it is useful to utilize a formula for 
a circulant determinant as in \cite{Nakayama1} 
\begin{align}
 \begin{vmatrix}
  f_1 & f_2 & f_3 & \cdots & f_k \\
  f_k & f_1 & f_2 & \cdots & f_{k-1} \\
  f_{k-1} & f_k & f_1 & \cdots & f_{k - 2} \\
  \vdots & \vdots & \vdots & \ddots & \vdots \\
  f_2 & f_3 & f_4 & \cdots & f_1   
 \end{vmatrix}
  = \prod_{I = 0} ^{ k - 1} (f_1 + \omega^{I} f_2 + \omega^{2 I} f_3 + 
                             \cdots + \omega^{(k-1) I} f_k )
 \label{cd}
\end{align}
with $\omega = \exp (2 \pi i /k)$. Using the identity
\begin{align}
 \prod_{I = 0}^{k - 1} [ \sum_{L = 0}^{k - 1} 
 ( t^{3 L} + t^{ 3 (k - L )} ) \omega^{L I} ]
  = \frac{(1 - t^{3 k})^k ( 1 - t^6)^k}{(1 - t^{3 k})^2} ~,
\end{align}
the index is computed as
\begin{align}
 {\cal I} = \prod_n  
 \frac{(1-t^{3 k n})^2 }
  {(1 - t^{2 n} v^n )^k(1-t^{2 n}/w^n)^k
   (1 - t^{2 n} w^n / v^n )^k}~. 
\end{align}

We would like to compare these indices with the one from the
gravity computation. However there is a subtle
problem whether we should remove the contribution
from the diagonal $U(1)^{k-1}$ part.
In the case of $AdS_5 \times S^5/{\mathbb Z}_k$, the
$U(1)^{k-1}$ part of the dual Yang-Mills theory is
removed to compare the indices with those from the
gravity computation \cite{Nakayama1}.
Suppose that the similar $U(1)$ decoupling should
be taken into account even in our case. 
The contribution from $U(1)$ part is just the same as that 
of $V=1$ case \eqref{indextriv}, so after the subtraction we have
\begin{align}
 {\cal I} = \prod_n  
 \frac{(1 - t^{3 k n})^2 (1 + t^{ 3 k n} )^{k-1}}
  {(1-t^{6n})^{k-1} (1 - t^{3 k n})^{k - 1}(1 - t^{2 n} v^n )(1-t^{2 n}/w^n) (1 - t^{2 n} w^n / v^n )}~. 
 \label{indexorb}
\end{align}
We would like to study the $U(1)$ problem furthermore
to examine whether this is indeed the case.

The dual geometry is the orbifold $AdS_5/{\mathbb Z}_k \times S^5$ 
with the excitation of localized string states at the fixed point,
and the different excitations correspond to different vacua
of the gauge theory.
The index for the supergravity on $AdS_5/{\mathbb Z}_k \times S^5$
can be computed by acting the orbifold projection to the index in
the covering space case \cite{KMMR,Nakayama1}.
The single-particle index can be computed as
\begin{align}
\begin{aligned}
 {\cal I} ' (t , v, w )
  & = \frac{1}{k} \sum_{I = 1}^k
    \left[ \frac{t^2 v}{1 - t^2 v} +  
    \frac{t^2 /w }{1 - t^2 /w } + \frac{t^2 w/v }{1 - t^2 w/v} 
    - \frac{t^3 \omega^I }{1 - t^3 \omega^I } 
    - \frac{t^3 / \omega^I }{1 - t^3 / \omega^I } \right] \\
  & =
    \frac{t^2 v}{1 - t^2 v} +  
    \frac{t^2 /w }{1 - t^2 /w } + \frac{t^2 w/v }{1 - t^2 w/v} 
    - \frac{2 t^{3 k} }{1 - t^{ 3 k} }  ~,
\end{aligned}
\end{align}
and the index including the multi-particle contribution is given as
\begin{align}
 {\cal I} = \exp \sum_{n = 1}^{\infty} 
   \frac{1}{n} {\cal I} ' (t^n , v^n , w^n )
  = \prod_{n = 1}^{\infty} \frac{(1-t^{3 k n})^2} 
  {(1 - t^{2 n} v^n) ( 1 - t^{2 n}/w^n ) (1 - t^{2 n} w^n /v^n )} ~.
\end{align}
The index is a slightly different form the both of 
\eqref{indextriv} and \eqref{indexorb}, and the difference
should be interpreted as the contribution from the 
localized string states.
It is desirable to include these stringy contributions
to the index from the AdS side and compare it with the
index of the gauge theory on the corresponding vacuum.

\baselineskip=11pt
\providecommand{\href}[2]{#2}\begingroup\raggedright\endgroup


\begin{thebibliography}{10}

\bibitem{Maldacena}
J.~M. Maldacena, ``The large {$N$} limit of superconformal field theories and
  supergravity,'' {\em Adv. Theor. Math. Phys.} {\bf 2} (1998) 231--252,
\href{http://www.arXiv.org/abs/hep-th/9711200}{{\tt hep-th/9711200}}.

\bibitem{Sundborg}
B.~Sundborg, ``The {Hagedorn} transition, deconfinement and {${ \cal N} = 4$}
  {SYM} theory,'' {\em Nucl. Phys.} {\bf B573} (2000) 349--363,
\href{http://www.arXiv.org/abs/hep-th/9908001}{{\tt hep-th/9908001}}.

\bibitem{AMMPR}
O.~Aharony, J.~Marsano, S.~Minwalla, K.~Papadodimas, and M.~Van~Raamsdonk,
  ``The {Hagedorn}/deconfinement phase transition in weakly coupled large {$N$}
  gauge theories,'' {\em Adv. Theor. Math. Phys.} {\bf 8} (2004) 603--696,
\href{http://www.arXiv.org/abs/hep-th/0310285}{{\tt hep-th/0310285}}.

\bibitem{HP}
S.~W. Hawking and D.~N. Page, ``Thermodynamics of black holes in {Anti-de
  Sitter} space,'' {\em Commun. Math. Phys.} {\bf 87} (1983)
577.

\bibitem{Wittent}
E.~Witten, ``Anti-de {Sitter} space, thermal phase transition, and confinement
  in gauge theories,'' {\em Adv. Theor. Math. Phys.} {\bf 2} (1998) 505--532,
\href{http://www.arXiv.org/abs/hep-th/9803131}{{\tt hep-th/9803131}}.

\bibitem{LM}
H.~Lin and J.~M. Maldacena, ``Fivebranes from gauge theory,''
\href{http://www.arXiv.org/abs/hep-th/0509235}{{\tt hep-th/0509235}}.

\bibitem{ITT}
G.~Ishiki, Y.~Takayama, and A.~Tsuchiya, ``{${\cal N} = 4$} {SYM} on {$R \times
  S^3$} and theories with 16 supercharges,'' {\em JHEP} {\bf 10} (2006) 007,
\href{http://www.arXiv.org/abs/hep-th/0605163}{{\tt hep-th/0605163}}.

\bibitem{ISTT}
G.~Ishiki, S.~Shimasaki, Y.~Takayama, and A.~Tsuchiya, ``Embedding of theories
  with {$SU(2|4)$} symmetry into the plane wave matrix model,''
\href{http://www.arXiv.org/abs/hep-th/0610038}{{\tt hep-th/0610038}}.

\bibitem{APS}
A.~Adams, J.~Polchinski, and E.~Silverstein, ``Don't panic! {Closed} string
  tachyons in {ALE} space-times,'' {\em JHEP} {\bf 10} (2001) 029,
\href{http://www.arXiv.org/abs/hep-th/0108075}{{\tt hep-th/0108075}}.

\bibitem{CM1}
R.~Clarkson and R.~B. Mann, ``{Eguchi-Hanson} solitons,''
\href{http://www.arXiv.org/abs/hep-th/0508109}{{\tt hep-th/0508109}}.

\bibitem{CM2}
R.~Clarkson and R.~B. Mann, ``{Eguchi-Hanson} solitons in odd dimensions,''
  {\em Class. Quant. Grav.} {\bf 23} (2006) 1507--1524,
\href{http://www.arXiv.org/abs/hep-th/0508200}{{\tt hep-th/0508200}}.

\bibitem{TZ}
A.~A. Tseytlin and K.~Zarembo, ``Effective potential in non-supersymmetric
  {$SU(N) \times SU(N)$} gauge theory and interactions of type 0 {D3-branes},''
  {\em Phys. Lett.} {\bf B457} (1999) 77--86,
\href{http://www.arXiv.org/abs/hep-th/9902095}{{\tt hep-th/9902095}}.

\bibitem{CST}
C.~Csaki, W.~Skiba, and J.~Terning, ``Beta functions of orbifold theories and
  the hierarchy problem,'' {\em Phys. Rev.} {\bf D61} (2000) 025019,
\href{http://www.arXiv.org/abs/hep-th/9906057}{{\tt hep-th/9906057}}.

\bibitem{AS}
A.~Adams and E.~Silverstein, ``Closed string tachyons, {AdS/CFT}, and large
  {$N$} {QCD},'' {\em Phys. Rev.} {\bf D64} (2001) 086001,
\href{http://www.arXiv.org/abs/hep-th/0103220}{{\tt hep-th/0103220}}.

\bibitem{DKR1}
A.~Dymarsky, I.~R. Klebanov, and R.~Roiban, ``Perturbative search for fixed
  lines in large {$N$} gauge theories,'' {\em JHEP} {\bf 08} (2005) 011,
\href{http://www.arXiv.org/abs/hep-th/0505099}{{\tt hep-th/0505099}}.

\bibitem{DKR2}
A.~Dymarsky, I.~R. Klebanov, and R.~Roiban, ``Perturbative gauge theory and
  closed string tachyons,'' {\em JHEP} {\bf 11} (2005) 038,
\href{http://www.arXiv.org/abs/hep-th/0509132}{{\tt hep-th/0509132}}.

\bibitem{GW}
D.~J. Gross and E.~Witten, ``Possible third order phase transition in the large
  {$N$} lattice gauge theory,'' {\em Phys. Rev.} {\bf D21} (1980)
446--453.

\bibitem{DM}
M.~R. Douglas and G.~W. Moore, ``D-branes, quivers, and {ALE} instantons,''
\href{http://www.arXiv.org/abs/hep-th/9603167}{{\tt hep-th/9603167}}.

\bibitem{KMMR}
J.~Kinney, J.~M. Maldacena, S.~Minwalla, and S.~Raju, ``An index for 4
  dimensional super conformal theories,''
\href{http://www.arXiv.org/abs/hep-th/0510251}{{\tt hep-th/0510251}}.

\bibitem{Nakayama1}
Y.~Nakayama, ``Index for orbifold quiver gauge theories,'' {\em Phys. Lett.}
  {\bf B636} (2006) 132--136,
\href{http://www.arXiv.org/abs/hep-th/0512280}{{\tt hep-th/0512280}}.

\bibitem{Cutkosky}
R.~E. Cutkosky, ``Harmonic functions and matrix elements for hyperspherical
  quantum field models,'' {\em J. Math. Phys.} {\bf 25} (1984)
939.

\bibitem{KS}
S.~Kachru and E.~Silverstein, ``4d conformal theories and strings on
  orbifolds,'' {\em Phys. Rev. Lett.} {\bf 80} (1998) 4855--4858,
\href{http://www.arXiv.org/abs/hep-th/9802183}{{\tt hep-th/9802183}}.

\bibitem{Liu}
H.~Liu, ``Fine structure of {Hagedorn} transitions,''
\href{http://www.arXiv.org/abs/hep-th/0408001}{{\tt hep-th/0408001}}.

\bibitem{ALW}
L.~Alvarez-Gaume, C.~Gomez, H.~Liu, and S.~Wadia, ``Finite temperature
  effective action, {$AdS_5$} black holes, and {$1/N$} expansion,'' {\em Phys.
  Rev.} {\bf D71} (2005) 124023,
\href{http://www.arXiv.org/abs/hep-th/0502227}{{\tt hep-th/0502227}}.

\bibitem{ABMW}
L.~Alvarez-Gaume, P.~Basu, M.~Marino, and S.~R. Wadia, ``Blackhole/string
  transition for the small {Schwarzschild} blackhole of {$AdS_5 \times S^5$}
  and critical unitary matrix models,''
\href{http://www.arXiv.org/abs/hep-th/0605041}{{\tt hep-th/0605041}}.

\bibitem{BD}
N.~D. Birrell and P.~C.~W. Davies, {\em Quantum fields in curved space}.
\newblock Cambridge University Press, 1982.

\bibitem{JZ}
J.~Jurkiewicz and K.~Zalewski, ``Vacuum structure of the {$U(N \to \infty)$}
  gauge theory on a two-dimensional lattice for a broad class of variant
  actions,'' {\em Nucl. Phys.} {\bf B220} (1983)
167.

\bibitem{HJ}
G.~T. Horowitz and T.~Jacobson, ``Note on gauge theories on {$M/\Gamma$} and
  the {AdS/CFT} correspondence,'' {\em JHEP} {\bf 01} (2002) 013,
\href{http://www.arXiv.org/abs/hep-th/0112131}{{\tt hep-th/0112131}}.

\bibitem{BK}
V.~Balasubramanian and P.~Kraus, ``A stress tensor for {Anti-de Sitter}
  gravity,'' {\em Commun. Math. Phys.} {\bf 208} (1999) 413--428,
\href{http://www.arXiv.org/abs/hep-th/9902121}{{\tt hep-th/9902121}}.

\bibitem{GKP}
S.~S. Gubser, I.~R. Klebanov, and A.~W. Peet, ``Entropy and temperature of
  black 3-branes,'' {\em Phys. Rev.} {\bf D54} (1996) 3915--3919,
\href{http://www.arXiv.org/abs/hep-th/9602135}{{\tt hep-th/9602135}}.

\bibitem{Rey}
S.-J. Rey and J.-T. Yee, ``Macroscopic strings as heavy quarks in large {$N$}
  gauge theory and {Anti-de Sitter} supergravity,'' {\em Eur. Phys. J.} {\bf
  C22} (2001) 379--394,
\href{http://www.arXiv.org/abs/hep-th/9803001}{{\tt hep-th/9803001}}.

\bibitem{Maldacenaw}
J.~M. Maldacena, ``Wilson loops in large {$N$} field theories,'' {\em Phys.
  Rev. Lett.} {\bf 80} (1998) 4859--4862,
\href{http://www.arXiv.org/abs/hep-th/9803002}{{\tt hep-th/9803002}}.

\bibitem{OV}
H.~Ooguri and C.~Vafa, ``Two-dimensional black hole and singularities of {CY}
  manifolds,'' {\em Nucl. Phys.} {\bf B463} (1996) 55--72,
\href{http://www.arXiv.org/abs/hep-th/9511164}{{\tt hep-th/9511164}}.

\bibitem{Horowitz}
G.~T. Horowitz, ``Tachyon condensation and black strings,'' {\em JHEP} {\bf 08}
  (2005) 091,
\href{http://www.arXiv.org/abs/hep-th/0506166}{{\tt hep-th/0506166}}.

\bibitem{HS}
G.~T. Horowitz and E.~Silverstein, ``The inside story: Quasilocal tachyons and
  black holes,'' {\em Phys. Rev.} {\bf D73} (2006) 064016,
\href{http://www.arXiv.org/abs/hep-th/0601032}{{\tt hep-th/0601032}}.

\bibitem{nuttier}
D.~Astefanesei, R.~B. Mann, and C.~Stelea, ``Nuttier bubbles,'' {\em JHEP} {\bf
  01} (2006) 043,
\href{http://www.arXiv.org/abs/hep-th/0508162}{{\tt hep-th/0508162}}.

\bibitem{GL1}
O.~Aharony, J.~Marsano, S.~Minwalla, and T.~Wiseman, ``Black hole-black string
  phase transitions in thermal 1+1 dimensional supersymmetric {Yang-Mills}
  theory on a circle,'' {\em Class. Quant. Grav.} {\bf 21} (2004) 5169--5192,
\href{http://www.arXiv.org/abs/hep-th/0406210}{{\tt hep-th/0406210}}.

\bibitem{GL2}
O.~Aharony, J.~Marsano, S.~Minwalla, K.~Papadodimas, M.~Van~Raamsdonk, and
  T.~Wiseman, ``The phase structure of low dimensional large {$N$} gauge
  theories on tori,'' {\em JHEP} {\bf 01} (2006) 140,
\href{http://www.arXiv.org/abs/hep-th/0508077}{{\tt hep-th/0508077}}.

\bibitem{GL}
R.~Gregory and R.~Laflamme, ``Black strings and p-branes are unstable,'' {\em
  Phys. Rev. Lett.} {\bf 70} (1993) 2837--2840,
\href{http://www.arXiv.org/abs/hep-th/9301052}{{\tt hep-th/9301052}}.

\bibitem{Schnitze}
H.~J. Schnitzer, ``Confinement/deconfinement transition of large {$N$} gauge
  theories with {$N_f$} fundamentals: {$N_f/N$} finite,'' {\em Nucl. Phys.}
  {\bf B695} (2004) 267--282,
\href{http://www.arXiv.org/abs/hep-th/0402219}{{\tt hep-th/0402219}}.

\bibitem{BMN}
D.~Berenstein, J.~M. Maldacena, and H.~Nastase, ``Strings in flat space and pp
  waves from {${\cal N} = 4$} super {Yang Mills},'' {\em JHEP} {\bf 04} (2002)
  013,
\href{http://www.arXiv.org/abs/hep-th/0202021}{{\tt hep-th/0202021}}.

\bibitem{FSS}
K.~Furuuchi, E.~Schreiber, and G.~W. Semenoff, ``Five-brane thermodynamics from
  the matrix model,''
\href{http://www.arXiv.org/abs/hep-th/0310286}{{\tt hep-th/0310286}}.

\bibitem{Semenoff}
G.~W. Semenoff, ``Matrix model thermodynamics,''
\href{http://www.arXiv.org/abs/hep-th/0405107}{{\tt hep-th/0405107}}.

\bibitem{SRV}
M.~Spradlin, M.~Van~Raamsdonk, and A.~Volovich, ``Two-loop partition function
  in the planar plane-wave matrix model,'' {\em Phys. Lett.} {\bf B603} (2004)
  239--248,
\href{http://www.arXiv.org/abs/hep-th/0409178}{{\tt hep-th/0409178}}.

\bibitem{HRSY}
S.~Hadizadeh, B.~Ramadanovic, G.~W. Semenoff, and D.~Young, ``Free energy and
  phase transition of the matrix model on a plane-wave,'' {\em Phys. Rev.} {\bf
  D71} (2005) 065016,
\href{http://www.arXiv.org/abs/hep-th/0409318}{{\tt hep-th/0409318}}.

\bibitem{KNY}
N.~Kawahara, J.~Nishimura, and K.~Yoshida, ``Dynamical aspects of the
  plane-wave matrix model at finite temperature,'' {\em JHEP} {\bf 06} (2006)
  052,
\href{http://www.arXiv.org/abs/hep-th/0601170}{{\tt hep-th/0601170}}.

\bibitem{Ling}
H.~Ling, A.~R. Mohazab, H.-H. Shieh, G.~van Anders, and M.~Van~Raamsdonk,
  ``Little string theory from a double-scaled matrix model,''
\href{http://www.arXiv.org/abs/hep-th/0606014}{{\tt hep-th/0606014}}.

\bibitem{Lin}
H.~Lin, ``Instantons, supersymmetric vacua, and emergent geometries,''
\href{http://www.arXiv.org/abs/hep-th/0609186}{{\tt hep-th/0609186}}.

\bibitem{AMMPR2}
O.~Aharony, J.~Marsano, S.~Minwalla, K.~Papadodimas, and M.~Van~Raamsdonk, ``A
  first order deconfinement transition in large {$N$} {Yang-Mills} theory on a
  small {$S^3$},'' {\em Phys. Rev.} {\bf D71} (2005) 125018,
\href{http://www.arXiv.org/abs/hep-th/0502149}{{\tt hep-th/0502149}}.

\bibitem{HO}
T.~Harmark and M.~Orselli, ``Quantum mechanical sectors in thermal {${\cal N} =
  4$} super {Yang-Mills} on {$R \times S^3$},''
\href{http://www.arXiv.org/abs/hep-th/0605234}{{\tt hep-th/0605234}}.

\bibitem{HO2}
T.~Harmark and M.~Orselli, ``Matching the {Hagedorn} temperature in
  {AdS/CFT},''
\href{http://www.arXiv.org/abs/hep-th/0608115}{{\tt hep-th/0608115}}.

\bibitem{Romelsberger}
C.~Romelsberger, ``An index to count chiral primaries in {$N = 1$} {$d = 4$}
  superconformal field theories,''
\href{http://www.arXiv.org/abs/hep-th/0510060}{{\tt hep-th/0510060}}.

\bibitem{Nakayama2}
Y.~Nakayama, ``Index for supergravity on {$AdS_5 \times T^{1,1}$} and conifold
  gauge theory,''
\href{http://www.arXiv.org/abs/hep-th/0602284}{{\tt hep-th/0602284}}.

\end{thebibliography}
\end{document}